# Optical Fourier surfaces

Nolan Lassaline[1], Raphael Brechbühler[1], Sander J. W. Vonk[1,2], Korneel Ridderbeek[1], Martin Spieser[3], Samuel Bisig[3], Boris le Feber[1], Freddy T. Rabouw[1,2], and David J. Norris[1]

[1]Optical Materials Engineering Lab, Dept. of Mechanical and Process Engineering, ETH Zurich, 8092 Zurich, Switzerland.
[2]Debye Institute for Nanomaterials Science, Utrecht University, Princetonplein 1, 3584 CC Utrecht, The Netherlands.
[3]Heidelberg Instruments Nano/SwissLitho, Technoparkstrasse 1, 8005 Zurich, Switzerland.

**Gratings[1] and holograms[2] are patterned surfaces that tailor optical signals by diffraction. Despite their long history, variants with remarkable functionalities continue to be discovered[3,4]. Further advances could exploit Fourier optics[5], which specifies the surface pattern that generates a desired diffracted output through its Fourier transform. To shape the optical wavefront, the ideal surface profile should contain a precise sum of sinusoidal waves, each with a well-defined amplitude, spatial frequency, and phase. However, because fabrication techniques typically yield profiles with at most a few depth levels, complex 'wavy' surfaces cannot be obtained, limiting the straightforward mathematical design and implementation of sophisticated diffractive optics. Here we present a simple yet powerful approach to eliminate this design–fabrication mismatch by demonstrating optical surfaces that contain an arbitrary number of specified sinusoids. We combine thermal scanning-probe lithography[6-8] and templating[9] to create periodic and aperiodic surface patterns with continuous depth control and sub-wavelength spatial resolution. Multicomponent linear gratings allow precise manipulation of electromagnetic signals through Fourier-spectrum engineering[10]. Consequently, we immediately resolve an important problem in photonics by creating a single-layer grating that simultaneously couples red, green, and blue light at the same angle of incidence. More broadly, we analytically design and accurately replicate intricate two-dimensional moiré patterns[11], quasicrystals[12,13], and holograms[14,15], demonstrating a variety of previously impossible diffractive surfaces. Therefore, this approach provides instant benefit for optical devices (biosensors[16], lasers[17,18], metasurfaces[4], and modulators[19]) and emerging topics in photonics (topological structures[20], transformation optics[21], and valleytronics[22]).**

Any patterned optical surface can be described as a Fourier sum of sinusoidal waves. Each component represents a specific spatial frequency ($g = 2\pi/\Lambda$ with period $\Lambda$) that interacts with an impinging beam. For applications, diffractive surfaces should ideally contain only the frequencies of interest. However, they are typically obtained by etching patterns into thin films to a fixed depth, creating arrays of vertical elements with shapes (trenches, holes, pillars) dictated by fabrication rather than design. These not only contribute unwanted spatial frequencies, complicating the optical response, but restrict the number of desired Fourier components that can be included. Clever placement of the elements (for example, aperiodically[10,12,13,17]) can offer some additional control. Alternatively, the collective response from arrays of smaller elements—nanoscale, sub-wavelength resonators—can be exploited in 'metasurfaces'[23]. However, no approach has yet offered complete control over the Fourier components in a diffractive surface. If available, simple analytical formulas could immediately specify the sum of sinusoids needed to obtain a complex desired output.

Wavy surfaces are in principle possible by grayscale lithography[24], which spatially adjusts the exposure of a polymeric resist to pattern with multiple depth levels. The profile can then be transferred into the underlying substrate via etching. However, grayscale lithography has not yet provided sufficient spatial resolution or depth control to create arbitrary optical surfaces. Similarly, interference lithography, which exposes the resist to multiple overlapping optical beams, can generate complex diffractive surfaces[25,26]. But they contain at most a few spatial frequencies, constrained by the exposure wavelengths.

To obtain arbitrary control over the Fourier components, we first design our structure by taking the Fourier transform of the desired diffraction pattern. After converting this analytical function into a two-dimensional (2D) grayscale bitmap (8-bit depth with 10×10 nm$^2$ pixels; see Methods and Extended Data Fig. 1), we then use thermal scanning-probe lithography[6-8] to raster-scan a heated cantilever with a sharp tip across a polymer film, locally removing material to match the bitmap depth at each pixel. Due to simultaneous monitoring of the surface topography by the tip for feedback, arbitrary surfaces with sub-nanometer depth control and high spatial resolution (<100 nm) can be written. These profiles can provide diffractive elements directly or be used as an etch mask or template. We exploit templating to replicate the pattern in other materials[9].

Figure 1 demonstrates our approach with one-dimensional (1D) sinusoidal gratings (periodic in $x$, constant in $y$), templated into Ag, with one, two, or three Fourier components (Fig. 1a,d,g). The insets show the targeted amplitudes, $A_i$, and spatial frequencies, $g_i$, for sinusoid $i$ (analytical formulas for all surfaces are in Methods). Because our structures are finite in size, their Fourier spectra will be slightly broadened from the analytical design (see modeling in Methods). The measured topographies for the patterns (Fig. 1b,e,h) show that the process faithfully reproduces the targeted profile with 1.8–2.3 nm root-mean-square (RMS) error (see Methods and Extended Data Fig. 2). These low values indicate that the desired Fourier components are predominant. Indeed, a detailed analysis for the single sinusoid (Extended Data Fig. 2) shows that the second harmonic dominates the error with an amplitude of only 3.5%· $A_1$ (0.9 nm).

To test the optical response of our gratings, we measure angle-resolved reflectivity spectra by imaging the back focal plane of an optical microscope onto a spectrometer (Methods and Extended Data Fig. 3a). Each sinusoidal component (here periodic in $x$) can provide momentum $\mathbf{g_i} = \frac{2\pi}{\Lambda_i}\hat{\mathbf{x}}$ (where $\hat{\mathbf{x}}$ is the unit vector along $x$) to an impinging beam. These contributions can affect the outgoing angle of the radiation or lead to electromagnetic surface waves—surface-plasmon polaritons (SPPs)—that propagate along the Ag–air interface with in-plane wavevector $\mathbf{k_{SPP}}$. We use the latter process (photon–SPP coupling) to characterize the capabilities of our surfaces.

We measure reflectivity as a function of in-plane wavevector $\mathbf{k_\parallel}$ of the incoming light. Figure 1c plots results for the single-sinusoidal grating for $\mathbf{k_\parallel} = k_x \hat{\mathbf{x}}$ (that is, energy *versus* $k_x$ with $k_y = 0$; see Extended Data Fig. 3b). Decreased reflectivity (orange lines) occurs when $\mathbf{k_\parallel} \pm \mathbf{g_1} = \mathbf{k_{SPP}}$. Thus, the grating creates a photon–SPP coupling channel, allowing the plasmonic dispersion to be optically probed. The match between the data and our analytical model (Fig. 1c; Methods), both here and below, confirms the fidelity of our process.

By including additional Fourier components, increasingly complex diffractive elements can be built. For two spatial frequencies $\mathbf{g_1}$ and $\mathbf{g_2}$ (Fig. 1d,e), two photon–SPP coupling channels open (Fig. 1f). Furthermore, SPP–SPP coupling arises if one of the spatial frequencies satisfies $\mathbf{k_{SPP}} \pm \mathbf{g}_i = \mathbf{k'_{SPP}}$, where $\mathbf{k_{SPP}}$, $\mathbf{k'_{SPP}}$ are wavevectors for SPPs propagating in different in-plane directions. This leads to a plasmonic stopband[27,28] (Extended Data Fig. 3b). Extended Data Fig. 4 shows an example at $\mathbf{k_\parallel} = \mathbf{0}$ when $\mathbf{g_2} = 2\mathbf{g_1}$. Although we have focused so far on the spatial frequencies of the sinusoids, our fabrication approach also allows independent control of their phase and amplitude. In Extended Data Fig. 4, phase is used to render either the upper or lower stopband edge 'dark' (not coupled to photons)[27]. Extended Data Fig. 5 uses amplitude to tune the stopband width (in energy) from 0 to ~0.5 eV. More generally, by adding further sinusoids, more complex plasmonic dispersions can be obtained. For example, Fig. 1g shows a three-component grating with multiple stopbands. These can be placed at arbitrary energies and incident photon angles. While the surface profile (Fig. 1h) would be difficult to intuit, Fourier design followed by our process leads directly to the desired response (Fig. 1i).

The control of sinusoidal components, shown above for 1D patterns with all $\mathbf{g}_i$ along $\hat{\mathbf{x}}$, can be extended to 2D (Extended Data Fig. 6). For example, if we sum two 1D sinusoids, one with $\mathbf{g_1}$ along $\hat{\mathbf{x}}$ and the other with $\mathbf{g_2}$ rotated by 10°, we obtain a moiré spatial interference pattern (Fig. 2a). If the rotation is 40°, the pattern in Fig. 2b results. Because these gratings now provide in-plane momentum along both $\hat{\mathbf{x}}$ and $\hat{\mathbf{y}}$, we report their optical response as reflectivity *versus* both in-plane wavevectors $k_x$ and $k_y$, taking a fixed-energy slice from the full dispersion diagram (Extended Data Fig. 3c). The experimentally accessible wavevectors for such a '$k$-space image' (due to our finite collection angle) are within the solid white circles in Fig. 2c,d. The measured reflectivity exhibits two pairs of orange arcs, each pair representing solutions to $\mathbf{k}_\parallel \pm \mathbf{g}_i = \mathbf{k}_{\text{SPP}}$ (Extended Data Fig. 3d). For both examples (Fig. 2a,b), when the 2D Fourier transform of the design is overlaid (Fig. 2c,d), the Fourier components $\pm\mathbf{g_1}$ and $\pm\mathbf{g_2}$ appear as orange spots outside the white circle and quantitatively explain the measured arcs. Even for only a 10° rotation, which leads to subtle intricacies in the surface pattern, the expected diffraction is observed.

Our approach can also exploit different basis functions. For example, Extended Data Fig. 7 shows a circular sinusoidal grating and a moiré interference pattern generated from two such gratings. Functions with varying local spatial frequencies can also be employed. Figure 2e shows a sinusoidal 'zone plate' (Methods), a fundamental element in Fourier optics[5]. It can act as a Fresnel lens to focus electromagnetic radiation by diffraction, representing a unit of holographic information. Currently, our process provides zone plates with dimensions appropriate for X-ray optics[29,30], with the added benefit of continuous depth control, highly desirable for this application[31]. Expanding our pattern to larger dimensions would lead to devices for ultraviolet or visible wavelengths. By superposition of many such patterns, holographic images with controlled Fourier spectra can be designed and implemented.

While the number of spatial components is arbitrary, several important symmetries can be generated with a finite number of sinusoids. Figure 3a,b shows a periodic pattern created from three 1D sinusoids with 60° rotation between them. The resulting profile is hexagonal, with 6-fold rotational symmetry, a common design for 2D arrays of holes or pillars. However, in our structure, the 2D Fourier spectrum is specified. The corresponding $k$-space image (Fig. 3c) reveals six orange arcs from photon–SPP coupling. Figure 3d,e shows a surface with 12-fold rotational symmetry created from six 1D sinusoids with 30° rotation between them. In $k$-space, 12 orange arcs appear (Fig. 3f). This profile, which does not possess translational symmetry, would be quasiperiodic if infinitely extended. Similar photonic quasicrystals using quasiperiodic arrays of trenches or holes have been reported for laser applications[10,17,32]. However, optimizing their design is computationally intensive and still results in 2D Fourier spectra with many unwanted spatial frequencies. Our structures are designed with simple analytical functions and exhibit precise control over the Fourier components.

To demonstrate the utility of our approach, we address a specific problem in photonics. Virtual- and augmented-reality hardware rely on optical systems for image generation and display[33]. The push for miniaturization has led to waveguide systems integrated in a single thin layer that exploit diffractive optics for in- and out-coupling of light[34]. For these devices, red, green, and blue photons should ideally be diffracted between free-space beams and propagating waveguide modes at a common angle. But current single-spatial-frequency gratings cause these colors to diffract at different, highly specific angles, resulting in bigger, more complicated devices. Solutions have been proposed, such as stacking three wavelength-specific diffractive layers[35], but all still retain disadvantages that prevent smaller, simplified systems.

With Fourier surfaces, a simple solution is immediately available. Three spatial frequencies can be included on a single surface to diffract all three colors at a common angle. Figure 4a,b shows such a profile, designed, implemented, and templated in Ag. The three 1D sinusoidal components simultaneously couple red, green, and blue photons at normal incidence (Fig. 4c), as seen by the three reflectivity dips in Fig. 4d, which arise due to photon–SPP coupling.

While we have focused on Ag structures, optical Fourier surfaces can be replicated in numerous materials. First, we have patterned (not shown) high-refractive-index polymers directly (Methods). Second, the polymer profile can be transferred into substrates via etching, for example Si (Fig. 4e) or $SiN_x$ (Extended Data Fig. 8). This also allows amplification of the profile depth[8]. Finally, either the patterned polymer or etched substrate can be used as a template[9]. For example, Extended Data Fig. 9 shows a $TiO_2$ Fourier surface templated from an etched Si substrate.

These results demonstrate precise fabrication of diffractive surfaces applicable to a broad spectral range (X-ray to infrared). Templating, extendable to rollable substrates[36], enables high-throughput production of many materials including active and multilayer solids[37,38]. In addition, diffractive surfaces can be accurately placed within or on top of elements in integrated photonic devices, allowing miniaturized optical systems[19,34]. Thus, researchers in photonics can exploit the previously unavailable capabilities of optical Fourier surfaces to address applications as well as explore emerging phenomena.

**Online Content** Methods, along with any Extended Data display items, are available in the online version of the paper; references unique to these sections appear only in the online paper.

**References**
1. Hopkinson, F. & Rittenhouse, D. An optical problem, proposed by Mr. Hopkinson, and solved by Mr. Rittenhouse. *Trans. Am. Philos. Soc.* **2**, 201-206 (1786).
2. Gabor, D. A new microscopic principle. *Nature* **161**, 777-778 (1948).
3. Ebbesen, T. W., Lezec, H. J., Ghaemi, H. F., Thio, T. & Wolff, P. A. Extraordinary optical transmission through sub-wavelength hole arrays. *Nature* **391**, 667-669 (1998).
4. Khorasaninejad, M., Chen, W. T., Devlin, R. C., Oh, J., Zhu, A. Y. & Capasso, F. Metalenses at visible wavelengths: diffraction-limited focusing and subwavelength resolution imaging. *Science* **352**, 1190-1194 (2016).
5. Goodman, J. W., *Introduction to Fourier Optics*. (W. H. Freeman, New York, 2017).

6. Mamin, H. J. & Rugar, D. Thermomechanical writing with an atomic force microscope tip. *Appl. Phys. Lett.* **61**, 1003-1005 (1992).
7. Pires, D., Hedrick, J. L., De Silva, A., Frommer, J., Gotsmann, B., Wolf, H., Despont, M., Duerig, U. & Knoll, A. W. Nanoscale three-dimensional patterning of molecular resists by scanning probes. *Science* **328**, 732-735 (2010).
8. Rawlings, C. D., Zientek, M., Spieser, M., Urbonas, D., Stöferle, T., Mahrt, R. F., Lisunova, Y., Brugger, J., Duerig, U. & Knoll, A. W. Control of the interaction strength of photonic molecules by nanometer precise 3D fabrication. *Sci. Rep.* **7**, 16502 (2017).
9. Nagpal, P., Lindquist, N. C., Oh, S. H. & Norris, D. J. Ultrasmooth patterned metals for plasmonics and metamaterials. *Science* **325**, 594-597 (2009).
10. Blanchard, R., Menzel, S., Pflügl, C., Diehl, L., Wang, C., Huang, Y., Ryou, J. H., Dupuis, R. D., Dal Negro, L. & Capasso, F. Gratings with an aperiodic basis: single-mode emission in multi-wavelength lasers. *New J. Phys.* **13**, 113023 (2011).
11. Sunku, S. S., Ni, G. X., Jiang, B. Y., Yoo, H., Sternbach, A., McLeod, A. S., Stauber, T., Xiong, L., Taniguchi, T., Watanabe, K., Kim, P., Fogler, M. M. & Basov, D. N. Photonic crystals for nano-light in moire graphene superlattices. *Science* **362**, 1153-1156 (2018).
12. Matsui, T., Agrawal, A., Nahata, A. & Vardeny, Z. V. Transmission resonances through aperiodic arrays of subwavelength apertures. *Nature* **446**, 517-521 (2007).
13. Martins, E. R., Li, J. T., Liu, Y. K., Depauw, V., Chen, Z. X., Zhou, J. Y. & Krauss, T. F. Deterministic quasi-random nanostructures for photon control. *Nat. Commun.* **4**, 2665 (2013).
14. Ozaki, M., Kato, J.-i. & Kawata, S. Surface-plasmon holography with white-light illumination. *Science* **332**, 218-220 (2011).
15. Zheng, G., Mühlenbernd, H., Kenney, M., Li, G., Zentgraf, T. & Zhang, S. Metasurface holograms reaching 80% efficiency. *Nat. Nanotechnol.* **10**, 308-312 (2015).
16. Tittl, A., Leitis, A., Liu, M. K., Yesilkoy, F., Choi, D. Y., Neshev, D. N., Kivshar, Y. S. & Altug, H. Imaging-based molecular barcoding with pixelated dielectric metasurfaces. *Science* **360**, 1105-1109 (2018).
17. Mahler, L., Tredicucci, A., Beltram, F., Walther, C., Faist, J., Beere, H. E., Ritchie, D. A. & Wiersma, D. S. Quasi-periodic distributed feedback laser. *Nat. Photon.* **4**, 165-169 (2010).
18. Yoshida, M., De Zoysa, M., Ishizaki, K., Tanaka, Y., Kawasaki, M., Hatsuda, R., Song, B., Gelleta, J. & Noda, S. Double-lattice photonic-crystal resonators enabling high-brightness semiconductor lasers with symmetric narrow-divergence beams. *Nat. Mater.* **18**, 121-128 (2019).
19. Ayata, M., Fedoryshyn, Y., Heni, W., Baeuerle, B., Josten, A., Zahner, M., Koch, U., Salamin, Y., Hoessbacher, C., Haffner, C., Elder, D. L., Dalton, L. R. & Leuthold, J. High-speed plasmonic modulator in a single metal layer. *Science* **358**, 630-632 (2017).
20. Lu, L., Joannopoulos, J. D. & Soljacic, M. Topological photonics. *Nat. Photon.* **8**, 821-829 (2014).
21. Pendry, J. B., Huidobro, P. A., Luo, Y. & Galiffi, E. Compacted dimensions and singular plasmonic surfaces. *Science* **358**, 915-917 (2017).
22. Hu, G., Hong, X., Wang, K., Wu, J., Xu, H.-X., Zhao, W., Liu, W., Zhang, S., Garcia-Vidal, F., Wang, B., Lu, P. & Qiu, C.-W. Coherent steering of nonlinear chiral valley photons with a synthetic Au–WS$_2$ metasurface. *Nat. Photon.* **13**, 467-472 (2019).
23. Genevet, P., Capasso, F., Aieta, F., Khorasaninejad, M. & Devlin, R. Recent advances in planar optics: from plasmonic to dielectric metasurfaces. *Optica* **4**, 139-152 (2017).
24. Kim, J., Joy, D. C. & Lee, S. Y. Controlling resist thickness and etch depth for fabrication of 3D structures in electron-beam grayscale lithography. *Microelectron. Eng.* **84**, 2859-2864 (2007).
25. Dakss, M. L., Kuhn, L., Heidrich, P. F. & Scott, B. A. Grating coupler for efficient excitation of optical guided waves in thin films. *Appl. Phys. Lett.* **16**, 523-525 (1970).
26. Campbell, M., Sharp, D. N., Harrison, M. T., Denning, R. G. & Turberfield, A. J. Fabrication of photonic crystals for the visible spectrum by holographic lithography. *Nature* **404**, 53-56 (2000).
27. Barnes, W. L., Preist, T. W., Kitson, S. C. & Sambles, J. R. Physical origin of photonic energy gaps in the propagation of surface plasmons on gratings. *Phys. Rev. B* **54**, 6227-6244 (1996).

28. Joannopoulos, J. D., Johnson, S. G., Winn, J. N. & Meade, R. D., *Photonic Crystals: Molding the Flow of Light*. (Princeton University Press, Princeton, 2008).
29. Chao, W., Harteneck, B. D., Liddle, J. A., Anderson, E. H. & Attwood, D. T. Soft X-ray microscopy at a spatial resolution better than 15 nm. *Nature* **435**, 1210-1213 (2005).
30. Wang, Y., Yun, W. & Jacobsen, C. Achromatic Fresnel optics for wideband extreme-ultraviolet and X-ray imaging. *Nature* **424**, 50-53 (2003).
31. Di Fabrizio, E., Romanato, F., Gentili, M., Cabrini, S., Kaulich, B., Susini, J. & Barrett, R. High-efficiency multilevel zone plates for keV X-rays. *Nature* **401**, 895-898 (1999).
32. Vitiello, M. S., Nobile, M., Ronzani, A., Tredicucci, A., Castellano, F., Talora, V., Li, L., Linfield, E. H. & Davies, A. G. Photonic quasi-crystal terahertz lasers. *Nat. Commun.* **5**, 5884 (2014).
33. Levola, T. Diffractive optics for virtual reality displays. *J. Soc. Inf. Display* **14**, 467-475 (2006).
34. Huang, Z. Q., Marks, D. L. & Smith, D. R. Out-of-plane computer-generated multicolor waveguide holography. *Optica* **6**, 119-124 (2019).
35. Mukawa, H., Akutsu, K., Matsumura, I., Nakano, S., Yoshida, T., Kuwahara, M. & Aiki, K. A full-color eyewear display using planar waveguides with reflection volume holograms. *J. Soc. Inf. Display* **17**, 185-193 (2009).
36. Yoo, D., Johnson, T. W., Cherukulappurath, S., Norris, D. J. & Oh, S. H. Template-stripped tunable plasmonic devices on stretchable and rollable substrates. *ACS Nano* **9**, 10647-10654 (2015).
37. Wuttig, M., Bhaskaran, H. & Taubner, T. Phase-change materials for non-volatile photonic applications. *Nat. Photon.* **11**, 465-476 (2017).
38. Shaltout, A. M., Shalaev, V. M. & Brongersma, M. L. Spatiotemporal light control with active metasurfaces. *Science* **364**, eaat3100 (2019).

**Acknowledgements** We thank S. Bonanni, U. Drechsler, F. Enz, T. Kulmala, and R. Stutz for technical assistance and D. Chelladurai, U. Dürig, R. Keitel, A. Knoll, M. Kohli, N. Rotenberg, D. Thureja, J. Winkler, and H. Wolf for helpful discussions. This project was funded by the European Research Council under the European Union's Seventh Framework Program (FP/2007-2013) / ERC Grant Agreement Nr. 339905 (QuaDoPS Advanced Grant). F.T.R. and S.J.W.V. acknowledge support from the Netherlands Organisation for Scientific Research (Rubicon-680-50-1509, Rubicon-680-50-1513, Gravitation Program "Multiscale Catalytic Energy Conversion", Veni-722.017.002, OCENW.KLEIN.008).

**Author Contributions** N.L., B.L.F., and D.J.N. conceived the project. N.L., R.B., and S.J.W.V. designed the Fourier surfaces with input from K.R., F.T.R., and D.J.N. N.L. patterned the polymer surfaces with assistance from K.R., M.S., and S.B. N.L. and R.B. fabricated the optical Fourier surfaces with assistance from K.R. and M.S. N.L. performed the characterization and topography analysis of the Fourier surface structures. N.L. performed the optical experiments with assistance from R.B. N.L., S.J.W.V., and F.T.R. analyzed the optical data with input from R.B. F.T.R. developed the analytical model. N.L. and D.J.N. wrote the manuscript with input from all authors.

**Author Information** The authors declare the following potential competing financial interests: S.B. is employed by Heidelberg Instruments Nano (previously SwissLitho AG), a provider of thermal scanning-probe lithography tools. At the time of his contribution, M.S. worked for SwissLitho AG. N.L., R.B., F.T.R., and D.J.N. have filed a patent application related to ideas in this work. Readers are welcome to comment on the online version of the paper. Correspondence and requests for materials should be addressed to D.J.N. (dnorris@ethz.ch).

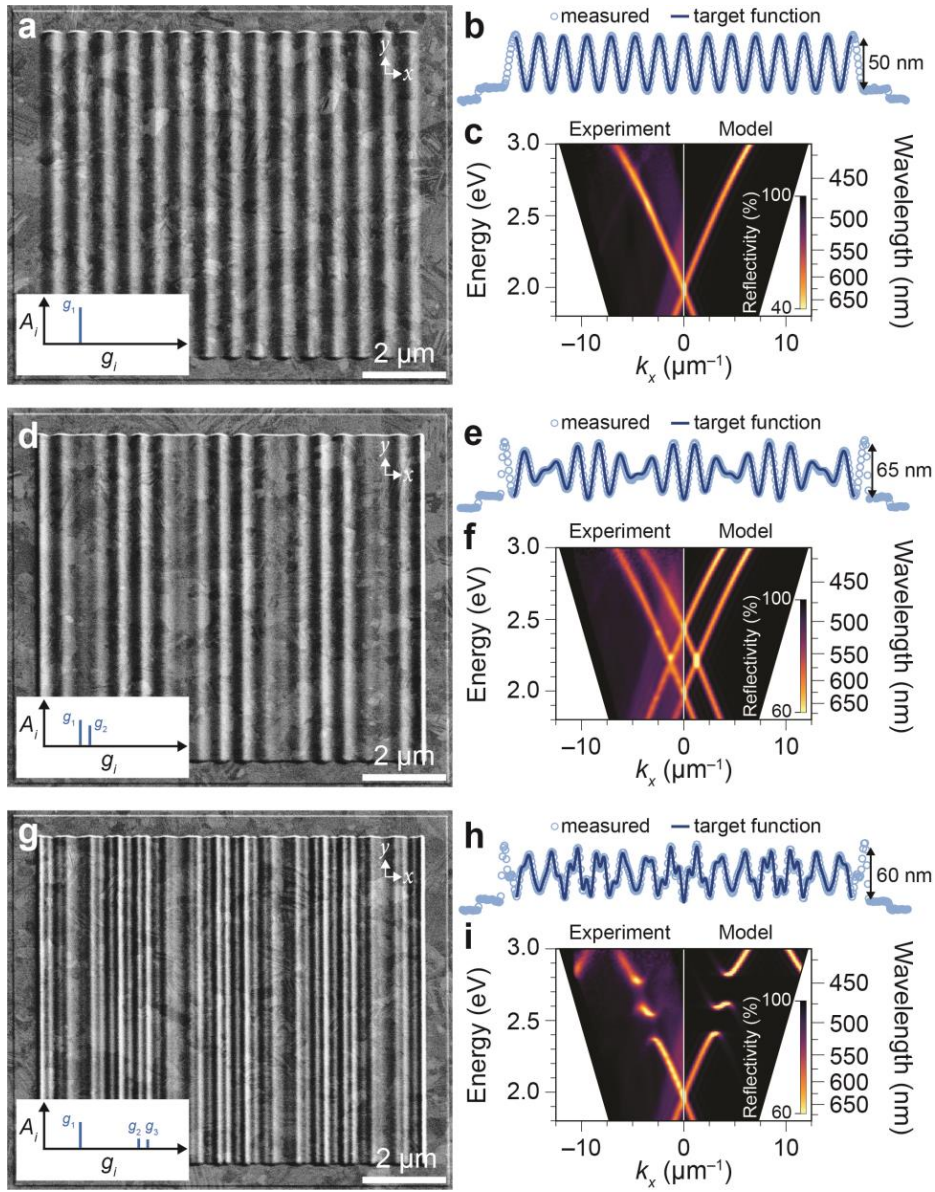

**Figure 1 | One-dimensional sinusoidal Fourier surfaces. a**,**d**,**g**, Scanning-electron micrographs (SEMs, 30° tilt) of Ag gratings with 1, 2, or 3 sinusoidal components. The insets show the sinusoidal amplitudes $A_i$ and spatial frequencies $g_i$. All design parameters are given in Extended Data Table 1. **b**,**e**,**h**, Measured (atomic force microscopy) and targeted surface topographies for the structures in **a**,**d**,**g**. Scan lengths are 11.3 µm. All target functions account for a slight distance miscalibration in the thermal scanning probe. The measured RMS error for the patterns are 1.8, 2.1, and 2.3 nm, respectively (see Methods). **c**, Experimental (left) and modeled (right) angle-resolved reflectivity spectra (energy *versus* in-plane photon wavevector along the grating, $k_x$, with $k_y \approx 0$) for the structure in **a**. The orange lines represent decreased reflectivity at photon angles that launch surface-plasmon polaritons (SPPs). These lines trace the SPP dispersion, displaced into the light cone by $g_1$. The black region represents energies and angles accessible in experiment (Extended Data Fig. 3). **f**, The two-component grating provides two photon–SPP coupling channels, doubling the orange lines. **i**, The three-component 1D sinusoidal Ag grating was designed to exhibit two plasmonic stopbands.

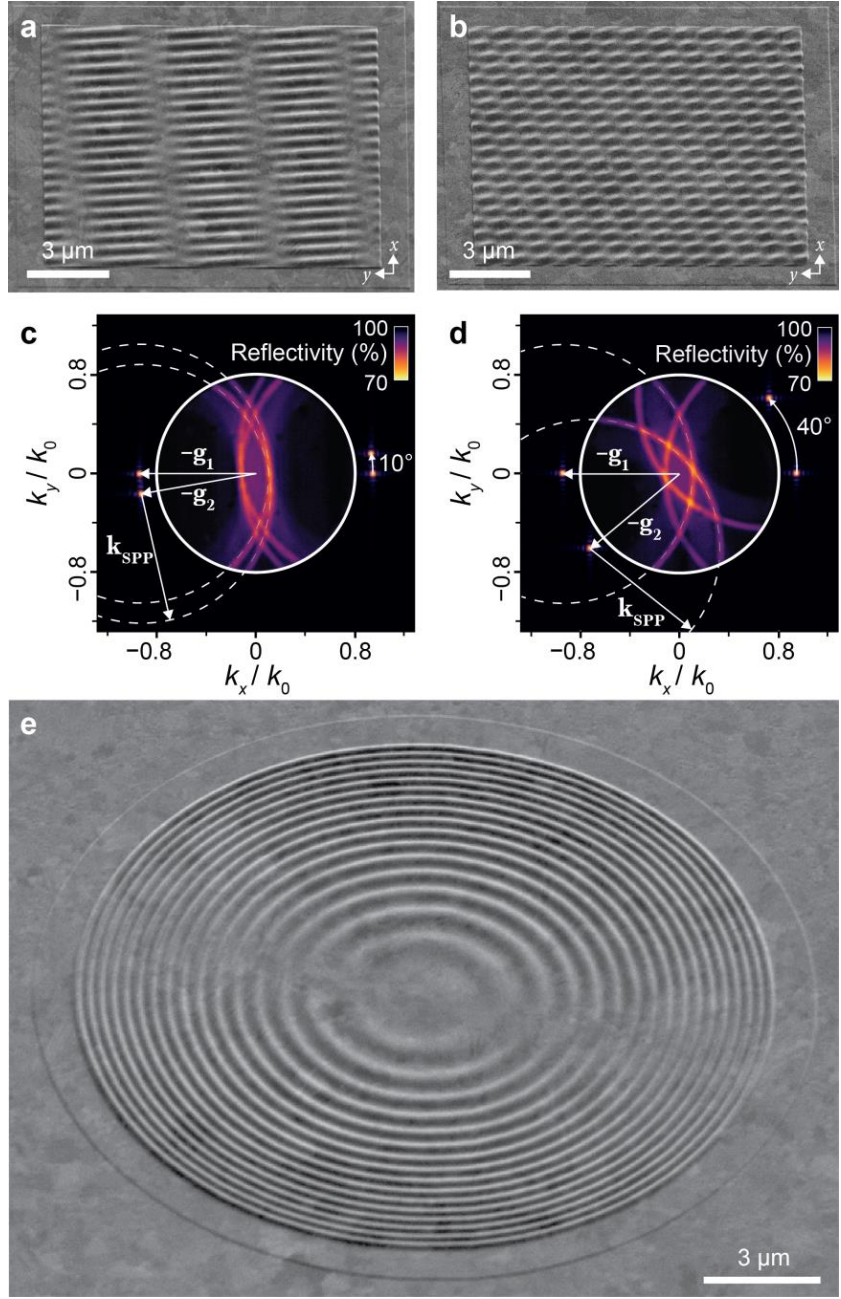

**Figure 2 | Two-dimensional Fourier surfaces. a,b**, SEMs (45° tilt) of moiré patterns in Ag from two superimposed 1D sinusoids: one with $\mathbf{g_1}$ along $\hat{\mathbf{x}}$ and the other with $\mathbf{g_2}$ rotated by 10° or 40°, respectively. See Extended Data Fig. 6. **c,d**, Measured $k$-space images (inside solid white circles) for photons (570 nm wavelength) reflected from patterns in **a,b**, respectively. $k_x$ and $k_y$ are normalized by the photon wavevector, $k_0$. Four orange arcs appear due to decreased reflectivity when photons launch surface-plasmon polaritons (SPPs) with wavevector $\mathbf{k_{SPP}}$, that is when $\mathbf{k_\parallel} \pm \mathbf{g_i} = \mathbf{k_{SPP}}$. $\pm\mathbf{g_1}$ and $\pm\mathbf{g_2}$ are shown as orange points outside the white circles. Their positions are determined from the 2D Fourier transform of the surface profiles used to define the structures. In **c,d**, we see that $\mathbf{k_\parallel} = -\mathbf{g_2} + \mathbf{k_{SPP}}$ forms an orange arc in $k$-space. **e**, SEM (45° tilt) of a Ag sinusoidal zone plate. For all structural design parameters, see Extended Data Table 1.

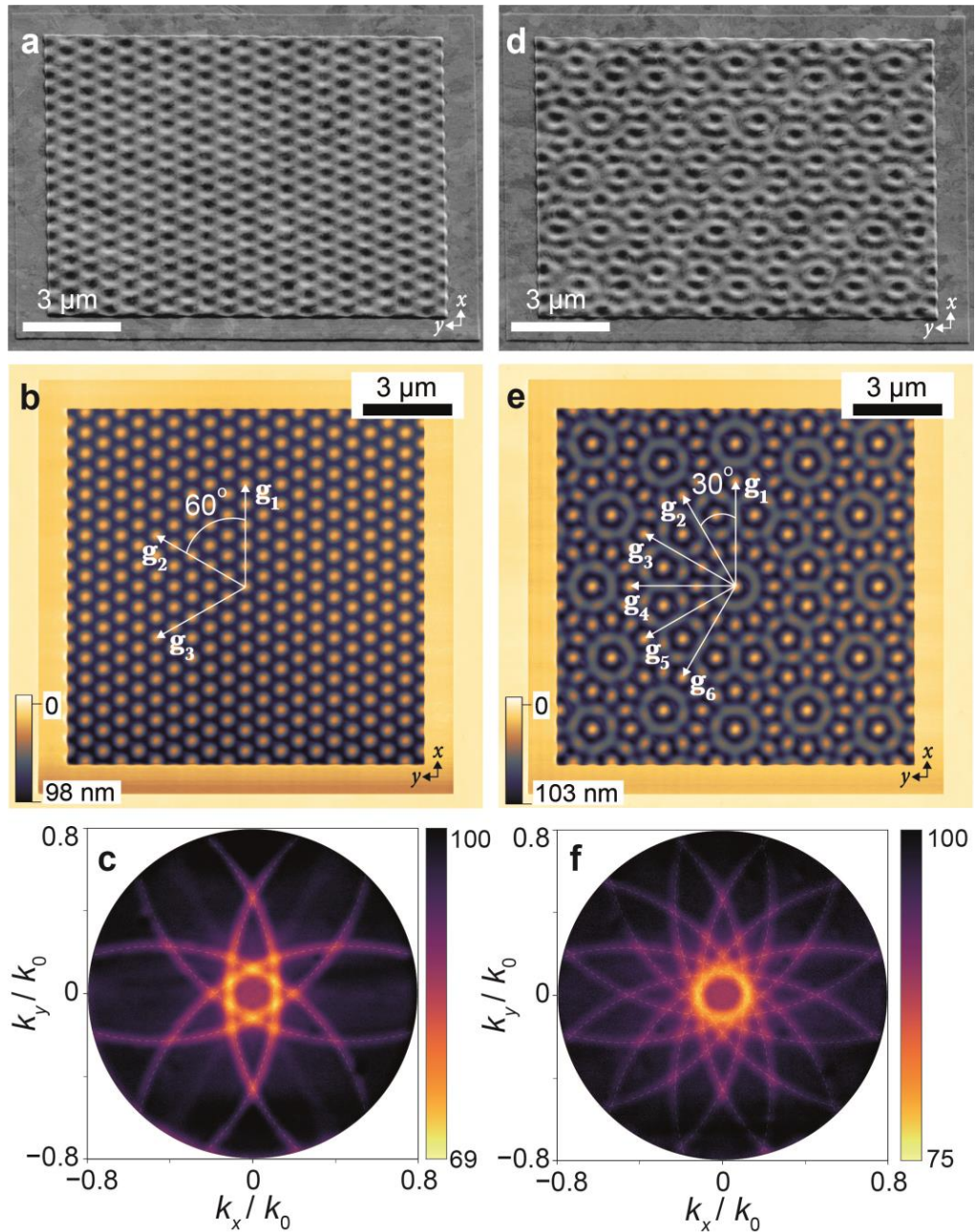

**Figure 3 | Periodic and quasiperiodic Fourier surfaces. a**,**d**, SEMs (45° tilt) of periodic and quasiperiodic optical Fourier surfaces templated in Ag with 6- and 12-fold rotational symmetry, defined when three and six 1D sinusoids are superimposed, respectively. **b**,**e**, Measured topographies (obtained during patterning) for the polymer films (PMMA/MA, see Methods) used to template the structures in **a**,**d**, respectively. All sinusoids have $\Lambda = 600$ nm and their corresponding vectors $\mathbf{g}_i$ are oriented in-plane, as shown, spaced by 60° and 30°, respectively. **c**,**f**, Measured $k$-space reflectivity images for photons (570 nm wavelength) incident on the patterns in **a**,**d**, respectively. 6 and 12 orange arcs appear due to decreased reflectivity when photons launch surface-plasmon polaritons (SPPs) with wavevector $k_{\mathrm{SPP}}$, that is, when $\mathbf{k}_\parallel \pm \mathbf{g}_i = \mathbf{k}_{\mathrm{SPP}}$ (dashed white lines). $k_x$ and $k_y$ are scaled by the photon wavevector, $k_0$. For all structural design parameters, see Extended Data Table 1.

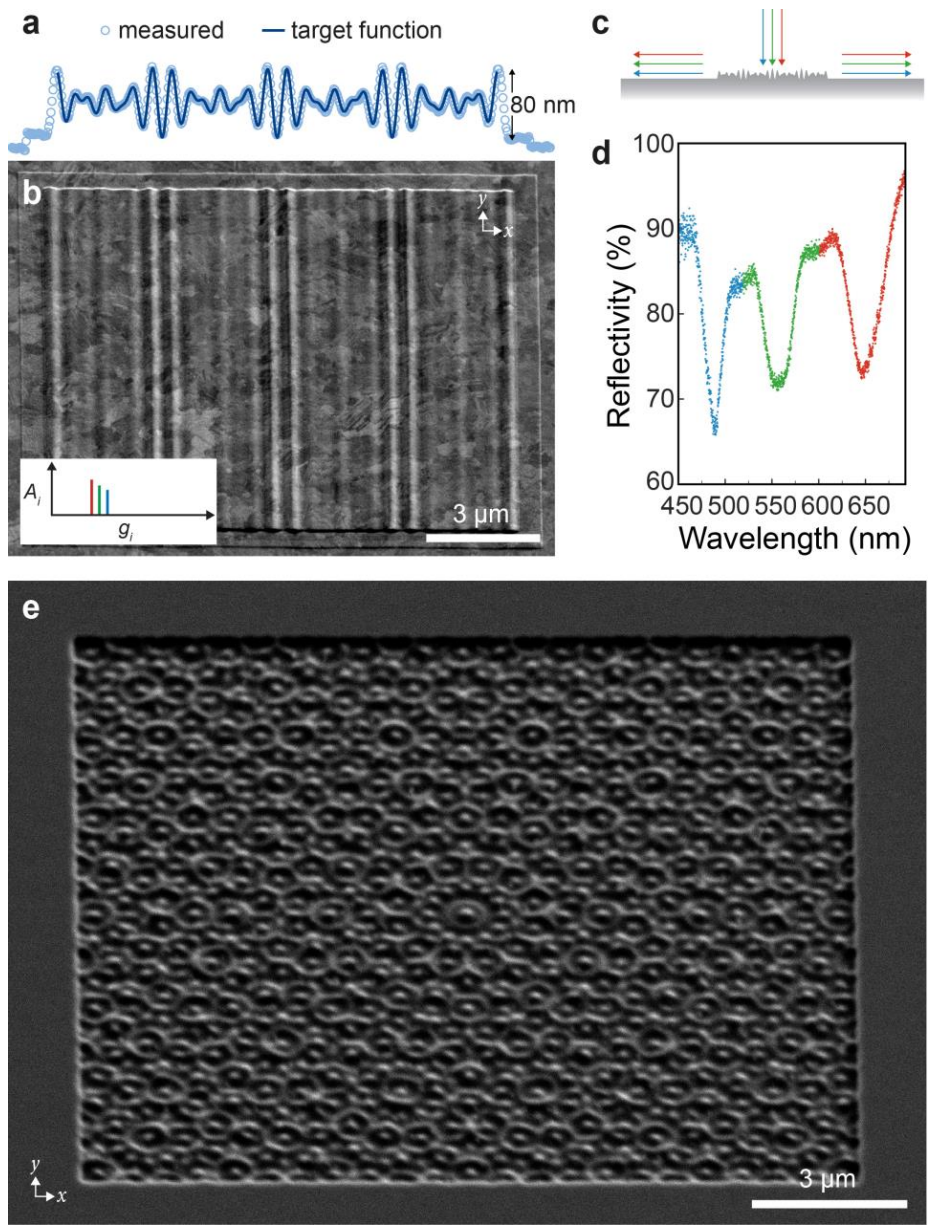

**Figure 4 | Applications of Fourier surfaces. a**, Comparison of the measured (atomic force microscopy) and targeted surface topography (accounting for a slight distance miscalibration in the thermal scanning probe) for a Ag Fourier surface that couples red, green, and blue photons at normal incidence to surface-plasmon polaritons. Scan length is 14.8 μm. The profile contains three 1D sinusoids with design periods $\Lambda_1 = 620$ nm, $\Lambda_2 = 520$ nm, and $\Lambda_3 = 445$ nm. **b**, SEM (45° tilt) of the Ag Fourier surface in **a**. The inset shows the sinusoidal amplitudes $A_i$ and spatial frequencies $g_i$. All design parameters are given in Extended Data Table 1. **c**, Cartoon of the coupling of red, green, and blue light simultaneously at normal incidence. **d**, Measured reflectivity as a function of photon wavelength for light at normal incidence (within ±1°). The three prominent reflectivity dips (colored as a visual guide) correspond to the coupling of red, green, and blue light at normal incidence. **e**, SEM (45° tilt) of a 12-fold rotationally symmetric quasicrystal, defined with twelve 1D sinusoids, etched into Si. For design parameters, see Extended Data Table 1.

## Methods

**Fourier surface design.** All surfaces were designed using analytical functions. In general, 1D real-space height profiles, $f(x)$, can be obtained from the desired Fourier spectrum, $F(k)$, via the 1D inverse Fourier transform,

$$f(x) = \frac{1}{2\pi} \int_{-\infty}^{\infty} F(k) e^{ikx} dk . \tag{1}$$

Similarly, 2D height profiles, $f(x, y)$, follow from the 2D inverse Fourier transform of $F(k_x, k_y)$,

$$f(x, y) = \frac{1}{(2\pi)^2} \iint_{-\infty}^{\infty} F(k_x, k_y) e^{i(k_x x + k_y y)} dk_x dk_y . \tag{2}$$

For $f(x)$ and $f(x, y)$, the origin is placed in the middle of the pattern for both $x$ and $y$. All functions are defined for the pattern in the polymer surface, where $x$ and $y$ lie in plane and $z$ is perpendicular. In these formulas, the height of the surface is defined relative to the unpatterned flat surface where $z = 0$. Note that the Fourier spectra in equations (1) and (2), used to calculate the infinitely-extended real-space surface profiles, neglect finite-size effects. The finite dimensions of the experimental profile lead to broadening of the Fourier spectra (see modeling in Methods).

For the 1D Fourier surfaces in Fig. 1, Extended Data Figs. 4, 5, and 8, and Fig. 4, the Fourier spectrum is sufficiently simple (with one, two, or three Fourier components, assuming infinite size in $x, y$) that the height profile can be written as a sum of sinusoids,

$$f(x) = \sum_i A_i \cos(g_i x + \varphi_i) - \Delta , \tag{3}$$

where $A_i$, $g_i$, and $\varphi_i$ are the amplitude, spatial frequency, and phase, respectively, for component $i$. Note that in equation (3), the sinusoidal surface profiles in the polymer are vertically shifted in $z$ by $-\Delta$. When templating is used to transfer the pattern to Ag, the surface profile is inverted and vertically shifted in $z$ by $+\Delta$. For clarity, all parameters for our polymer surfaces are provided in Extended Data Table 1.

For the 2D Fourier surfaces in Fig. 2a,b, Fig. 3a,d, Fig. 4e, and Extended Data Fig. 9, the height profile was given by

$$f(x,y) = \sum_i A_i \cos[g_i(x \cos \theta_i + y \sin \theta_i) + \varphi_i] - \Delta, \tag{4}$$

where $\theta_i$ is the in-plane rotation angle from the $x$ axis for component $i$. The 2D circular Fourier surfaces in Extended Data Fig. 7 follow

$$f(r,\theta) = \sum_i A_i \cos(g_i|\mathbf{r} - \mathbf{r}_i| + \varphi_i) - \Delta, \tag{5}$$

where $r$ and $\theta$ are the radial distance and polar angle, respectively. $\mathbf{r}$ is the coordinate in the surface plane and is a function of $r$ and $\theta$. $\mathbf{r}_i$ is the origin for circular component $i$. The sinusoidal zone plate[39] in Fig. 2e follows the function:

$$f(r) = A \sin\left[\pi \left(\frac{r}{L}\right)^2\right] - \Delta \tag{6}$$

where $A$ is an amplitude, and $L$ is a characteristic length.

**Bitmap generation.** The analytical functions are then converted into bitmaps. The overall dimensions in $x$ and $y$ are chosen for the structure, and the analytical function is mapped onto a 10×10 nm² pixel grid. The normalized depth of the structure in $z$ was assigned for each pixel by discretizing the total normalized depth to 256 levels (8-bit precision). The physical patterning depth was assigned for each pixel by inputting the maximum physical depth of the structure to the thermal scanning probe control software (see sample fabrication section below), which then assigned the physical depth for each pixel based on its 8-bit depth level. The entire process flow, from analytical mathematical design to pattern transfer to an optical material, is depicted in Extended Data Fig 1.

**Materials.** 1-mm-thick glass microscope slides and 1-mm-thick, 2-inch-diameter Si(100) wafers (1–10 Ωcm resistivity) were purchased from Paul Marienfeld and Silicon Materials, respectively. Ag (1/4-inch-diameter × 1/4-inch-long pellets, 99.999%), Au (1/8-inch-diameter × 1/8-inch-long pellets, 99.999%), TiO$_2$ sputter targets (200-mm diameter, 99.95%), and ultraviolet-curable (UV) epoxy (OG142-95 and OG116-31) were obtained from Kurt J. Lesker, ACI Alloys, FHR Anlagenbau, and Epoxy Technology, respectively. Tungsten dimple boats (49×12×0.4 mm$^3$) were bought from Umicore. Two polymer resists from Allresist GmbH were used: PMMA/MA [AR-P 617, poly(methyl methacrylate-co-methacrylic acid), 33% copolymer, 3% dilution in anisol] and CSAR [AR-P 6200, containing poly(α-methylstyrene-co-methyl chloroacrylate) in anisol]. Silicon cantilevers for thermal scanning-probe lithography with a tip radius of ~3-5 nm were provided by SwissLitho (SL2015-2-HPL, SL2016-3-HPL, SL2018-13-HPL, and SL2018-2-MBS). Hydrochloric (HCl, 37%) and nitric (HNO$_3$, ≥65%) acids were purchased from Sigma-Aldrich.

**Sample fabrication.** A Si wafer was typically used as the sample substrate. It was removed from its factory packaging in the cleanroom and used directly. The polymer resist layer was spin-coated onto it using a two-step procedure. For PMMA/MA or CSAR, the resist solution was deposited on the sample surface and accelerated at 500 r.p.m. s$^{-1}$ to 500 r.p.m. for 5 s. Then the PMMA/MA (CSAR) was accelerated at 2000 r.p.m. s$^{-1}$ to 2000 r.p.m. (2500 r.p.m.) for a total time of 40 s. After spin-coating, the PMMA/MA (CSAR) layer was baked at 180 °C for 5 min (150 °C for 1 min).

For thermal scanning-probe lithography, the substrate/polymer stack was placed in a NanoFrazor Explore (SwissLitho). A cantilever with a sharp tip was loaded into the cantilever holder, which was then attached to the NanoFrazor scan head. The tip was brought close to the sample and an auto-approach function achieved surface contact. The tip position, temperature response, and sample tilt were calibrated. The temperature at the base of the tip was set to an initial value between 700–950 °C, depending on the cantilever model. Calibration scans were performed to optimize the patterning depth of the sinusoidal structures. The bitmap defining the desired Fourier surface was then loaded into the NanoFrazor software. The tip was scanned across the patterning surface on a 10×10 nm$^2$ pixel grid. A force pulse (~6 μs) was applied at each pixel to match the depth level of the bitmap in the polymer resist. As the tip patterned the surface, it simultaneously measured the topography as in contact-mode atomic force microscopy (AFM). The measured error between the written pattern and the desired pattern was passed to a feedback loop such that the write force could be adjusted to reach the desired depth level, if necessary. The scan progressed until all pixels in the design were patterned into the surface, at which point the tip was available to write the next pattern.

To obtain Ag diffractive surfaces, Ag was thermally evaporated[40] (Kurt J. Lesker, Nano36) onto the patterned polymer film at a pressure of ~3×10$^{-7}$ mbar. A tungsten boat loaded with Ag pellets was heated to deposit at a rate of 25 Å s$^{-1}$. The process was stopped when the film thickness was ~750 nm. A glass slide was then affixed with UV-curable epoxy (OG142-95) onto the exposed Ag surface, and the glass/epoxy/Ag stack peeled off, revealing a Ag surface with the negative of the initial pattern in the polymer surface.

SiN$_X$ surfaces were obtained by using a Si/SiO$_2$/SiN$_x$ stack as a substrate. 2000 nm SiO$_2$ was thermally grown onto a Si wafer, followed by chemical vapor deposition of 200 nm of SiN$_X$. The wafer was diced into 1.5×1.5 cm$^2$ pieces for thermal scanning-probe lithography using PMMA/MA as the polymer. The pattern in the polymer film was transferred into the underlying SiN$_x$ substrate via reactive-ion etching (RIE, Oxford Instruments, NPG 80) using a gas mixture of 50 sccm CHF$_3$ and 5 sccm O$_2$. The etching was performed at a chamber pressure of 55 mTorr, with 100 W RF power and a SiN$_x$ etch rate of 45 nm min$^{-1}$ for 5 min, where the depth of the transferred pattern in SiN$_x$ was approximately the same as the depth in the polymer pattern (~1:1 selectivity). Afterwards, the substrate was ultrasonicated in acetone, followed by isopropanol, and blown dry with N$_2$.

To obtain Si surfaces for either direct use or for templating, the pattern in the polymer film was transferred into the underlying Si substrate via inductively coupled plasma (ICP) etching (Oxford Instruments, Plasma Pro) using a gas mixture of 17.0 sccm $SF_6$, 17.5 sccm $C_4F_8$, and 60 sccm Ar. The Si etching was done at a chamber pressure of 20 mTorr, with a forward power of 50 W, and at a rate of ~25 nm min$^{-1}$ for 6.33 min, where the depth of the transferred pattern in Si was approximately the same as the depth in the polymer pattern (~1:1 selectivity). After etching, the sample was sonicated for 2 min in acetone and 2 min in IPA, followed by 5 min of $O_2$ plasma cleaning at 600 W.

Patterned $TiO_2$ samples were obtained by using patterned Si templates. A 25-nm-thick Au layer was thermally evaporated onto the patterned Si wafer at a pressure of ~3×10$^{-7}$ mbar and a rate of 10 Å s$^{-1}$. $TiO_2$ was then RF sputtered onto the exposed gold surface (von Ardenne, CS 320 S) with 400 W, a chamber pressure of 4 ×10$^{-3}$ mbar, and 14 sccm Ar, for 160 min, resulting in a ~300-nm-thick film. A glass slide was then affixed with UV-curable epoxy (OG116-31) onto the exposed $TiO_2$ layer, and the glass/epoxy/$TiO_2$/Au stack peeled off, revealing a $TiO_2$/Au surface with the negative of the initial pattern in the Si surface. Finally, the Au layer was removed by immersing the sample in aqua regia (4:1 mixture of HCl:$HNO_3$) for 5 min. Afterwards, the sample was rinsed in deionized water and blown dry with $N_2$.

**Surface-topography characterization.** The topography of the Fourier surfaces was measured by the scanning probe during patterning and independently verified with AFM on the templated Ag surface. The topography of our Ag single-sinusoidal surface (Fig. 1a,b) is analyzed in Extended Data Fig. 2. AFM scans (Bruker, Dimension FastScan AFM with a Bruker NCHV-A cantilever) were collected in tapping mode under ambient conditions. The raw data was processed by first removing the instrumental high-frequency scan noise in the scanning-probe analysis software Gwyddion. Next, row alignment and plane-levelling were performed in MATLAB to obtain the corrected data, shown in Extended Data Fig. 2a. These data were then analyzed by fitting a 2D sinusoidal function (with the form shown in Extended Data Table 1 for Fig. 1a; periodic along $x$, constant in $y$), where the fit parameters and residuals were extracted. The amplitude and period of the fitted function was $A_1 = 25.5$ nm (2% larger than design value) and $\Lambda = 610$ nm (1.7% larger than design value), respectively. As we obtained a consistent horizontal distance error in both our etched and templated gratings, we attributed this error to a distance miscalibration in the thermal scanning probe. The RMS error between the 2D design function and measured topography for the structure in Fig. 1a was found to be 1.8 nm after this error was taken into account. A similar procedure was used to extract RMS errors for other Fourier surfaces, as reported in the figure captions. See Extended Data Fig. 2 for further details.

**Optical characterization.** The optical-characterization setup is depicted in Extended Data Fig. 3a. Ag surfaces were measured with an inverted optical microscope (Nikon, Eclipse Ti-U) equipped with a 50× air objective [Nikon, TU Plan Fluor, numerical aperture (NA) of 0.8]. A halogen lamp was used to illuminate the sample. The lamp filament was imaged to fill the back focal plane of the microscope objective. After a beamsplitter, the light was focused onto the sample and then collected by the same objective. Reflected light was transmitted through the beamsplitter and passed through a circular aperture in the real-space image plane to isolate the structure of interest. The back focal plane was projected onto the entrance slit of an imaging spectrograph (Andor Shamrock 303i) where it was relayed to a sensitive digital camera (Andor Zyla PLUS sCMOS) for image acquisition. Reflectivity measurements were obtained for both dispersed $k$-space measurements (Fig. 1c,f,i, Extended Data Figs. 4b and 5b–i) and $k$-space images (Figs. 2c,d and 3c,f), by acquiring a background image, a reference image, and a signal image. The background, reference, and signal images were recorded by acquiring the counts when no light was incident on the camera, when light was reflected from flat Ag on the sample, and when light was reflected from the pattern of interest, respectively. The final reflectivity image (% reflectivity) was calculated using:

$$\% \text{ reflectivity} = 100 \times \frac{\text{signal}-\text{background}}{\text{reference}-\text{background}} \qquad (7)$$

For the dispersed $k$-space measurements, a grating (150 lines mm$^{-1}$ blazed at 500 nm) was inserted into the imaging path in the spectrometer such that the light was spectrally dispersed along one axis of the camera. The spectrometer slit was parallel to $k_x$. A linear polarizer was inserted into the collection path to select only p-polarized light, which couples to SPPs. Thus, in a single acquisition, the dispersion relation (energy *versus* in-plane momentum along the grating, $k_x$, with $k_y \approx 0$) could be measured. The experimental window is overlaid with a schematic of the theoretical SPP dispersion in Extended Data Fig. 3b.

For the $k$-space images, a bandpass filter centered at 570 nm with a full-width-at-half-maximum (FWHM) of 10 nm was placed in the excitation path. The linear polarizer was removed from the collection path such that the measurement probed all polarizations equally. The slit at the entrance of the imaging spectrograph was opened completely and the $k$-space image was relayed to the camera using a mirror instead of a diffraction grating to eliminate stray diffracted light. A schematic of this measurement, performed at a narrow range of photon energies selected by the bandpass filter, is depicted in Extended Data Fig. 3d. A cartoon of the complete light cone and SPP dispersion is depicted in Extended Data Fig. 3c.

The reflectivity spectrum in Fig. 4d was obtained by plotting the dispersed $k$-space measurement for the three-component Fourier surface in Fig. 4b at a fixed angle of incidence (near normal incidence). Spectra were averaged over a collection angle of ±1°.

**Analytical model.** Optical modes bound to a periodic surface have an electric-field profile of the form

$$\mathbf{E_k}(\mathbf{r}) = e^{-i\mathbf{k}\cdot\mathbf{r}}\mathbf{u_k}(\mathbf{r}), \tag{8}$$

where **k** is the Bloch wavevector of the mode, and $\mathbf{u_k}(\mathbf{r})$ is a function with the same periodicity as the surface. We consider an arbitrary 1D grating profile, like those in Fig. 1 of the main text, for which all surface Fourier components $i$ have an in-plane wavevector $\mathbf{g}_i = g_i\hat{\mathbf{x}}$. The overall periodicity $2\pi/G$ of the surface profile can be much longer than any of the periodicities $\{2\pi/g_1, 2\pi/g_2, \ldots, 2\pi/g_N\}$ of the $N$ constituent sinusoids:

$$G^{-1} = \mathrm{lcm}(g_1^{-1}, g_2^{-1}, \ldots, g_N^{-1}), \tag{9}$$

where lcm denotes the least common multiple. For example, the grating in Fig. 1g of the main text has an overall design periodicity of $2\pi/G = 96.6$ μm and $G = 0.0650$ μm$^{-1}$. The full field profile of a mode $\mathbf{E_k}(\mathbf{r})$ contains all in-plane wavevector components $(k_x + nG, k_y)$ with any integer $n$. However, to calculate the plasmonic dispersion and stopbands of our 1D Fourier surfaces, we do not need the full field profile. Instead, we can use a relatively simple coupled-mode model with a limited basis, which only accounts for first-order coupling between plane waves differing in wavevector by $\mathbf{g}_i$ of one of the sinusoids of the grating.

On a flat Ag–dielectric interface, SPP modes have in-plane wavevector $\mathbf{k_{SPP}}$ with magnitude

$$k_{\mathrm{SPP}} = \frac{\omega}{c}\sqrt{\frac{\varepsilon_{\mathrm{m}}(\omega)\varepsilon_{\mathrm{d}}}{\varepsilon_{\mathrm{m}}(\omega)+\varepsilon_{\mathrm{d}}}}, \tag{10}$$

where $\omega$ is the SPP angular frequency, $c$ the speed of light, and $\varepsilon_{\mathrm{m}}$ is the frequency-dependent relative permittivity of the metal. The relative permittivity of the dielectric $\varepsilon_{\mathrm{d}}$ is assumed to be frequency independent. We note that when calculating $\mathbf{k_{SPP}}$ for Figs. 2 and 3 of the main text, we used $\varepsilon_{\mathrm{d}} = 1.061$. This value was determined by fitting the SPP dispersion for an independent sample. Extracting a permittivity slightly above 1 was perhaps due to residual polymer on the Ag surface after templating. For the structures in Fig. 1 of the main text, our fabrication process had been improved and $\varepsilon_d = 1$ was extracted and used for modeling.

In Fig. 1 of the main text, we measure the dispersion of our Fourier surfaces along the $k_x$-direction. Stopbands in this direction occur whenever $2k_{SPP} = g_i$ for one of the sinusoids $i$ in the grating. This occurs at energies

$$\hbar\omega_i = \frac{hc}{2n_{eff}\Lambda_i}, \tag{11}$$

where $\hbar = h/2\pi$ with $h$ as Planck's constant, and $n_{eff} = \sqrt{\varepsilon_m(\omega)\varepsilon_d/[\varepsilon_m(\omega)+\varepsilon_d]}$ is the effective refractive index of the SPP mode on the flat Ag–dielectric interface. While the SPP dispersion and any stopbands therein lie outside the light cone, we can measure a stopband if some sinusoid $j$ provides momentum to couple free-space photons to SPPs. The stopband will then appear in our reflectivity measurement at a photon in-plane wavevector equal to

$$k_{ij} = n_{eff}\frac{\omega_i}{c} - g_j = 2\pi\left(\frac{1}{2\Lambda_i} - \frac{1}{\Lambda_j}\right). \tag{12}$$

For a more rigorous calculation of the SPP dispersion for our 1D Fourier surfaces, we use a coupled-mode model. We couple plane waves with wavevector component $k_{x,0} = k_{SPP}$ to those with $k_{x,i} = k_{SPP} - g_i$ for all sinusoids $i \in \{1,2,...,N\}$ in the surface profile. The energies of the coupled modes are the eigenvalues of the interaction matrix $H$, which has dimensions $(N + 1) \times (N + 1)$. The diagonal elements of the matrix are the energies that a plane wave of wavevector $k_{x,i}$ would have on the flat Ag–dielectric interface, which we obtain by evaluating the inverse of equation (10), $\omega(k_{SPP})$, at $k_{SPP} = |k_{x,i}|$:

$$H_{ii} = \hbar\omega(|k_{x,i}|). \tag{13}$$

For this, we use the permittivity data $\varepsilon_m(\omega)$ of template-stripped Ag[39] and $\varepsilon_d = 1$ for air. The off-diagonal elements $H_{ij}$ describe the interaction between plane waves $i$ and $j$. We only consider coupling involving the fundamental SPP wave with wavevector $k_{x,0} = k_{\text{SPP}}$ and neglect coupling between plane waves $i \geq 1$ and $j \geq 1$:

$$H_{0i} = H_{i0} = \hbar \Gamma_i \ . \tag{14}$$

Here $\Gamma_i$ is the (real-valued) rate at which the surface sinusoid $i$ couples a plane wave with $k_{x,0}$ and a plane wave with $k_{x,i}$. This rate determines the width of the stopband $\Delta E_i \approx 2\hbar\Gamma_i$ due to grating component $i$. Extended Data Fig. 5 shows that we can control this by tuning the corresponding amplitude $A_i$ of the sinusoid[27]. For Fig. 1i in the main text, we estimated values of $\Gamma_i$ based on the dispersion data and plugged them into the model.

The eigenvalues of $H$ are the energies $E_i$ of the coupled modes, while the eigenvectors $\mathbf{v}_i$ describe their composition in terms of the plane wave basis functions. For each mode, the first component of the eigenvector $v_{i,0}$ is the SPP character of the coupled mode. In addition to coupling plane waves, each sinusoid $i$ of the surface profile can provide in-plane momentum for SPP excitation by free-space photons if $|k_{\text{SPP}} - g_i| \leq \omega/c$. Free-space photons with in-plane wavevector component $\mathbf{k}_{\parallel} = (k_x, 0)$ can excite SPPs if the momentum is matched through some sinusoid $i$ to a coupled mode $j$ with significant SPP character $v_{j,0}$.

For the calculated dispersion plots in Fig. 1 of the main text, we first solved the eigenvalues and eigenvectors of $H$ for a regular grid of $k_{\text{SPP}}$-values ranging from 0 and 25 μm$^{-1}$ in $M = 5001$ steps of 0.005 μm$^{-1}$. This yields, for each value $k_{\text{SPP},i}$, a set of $(N + 1)$ mode energies $E_{i,j}$ and $(N + 1)$ values $v_{i,j,0}$ for the corresponding SPP character (the coefficient for the fundamental wavevector component $k_{x,0} = k_{\text{SPP}}$ to the eigenvector of coupled mode $j$). In addition to coupling plane waves, each sinusoid $l$ of the surface profile can provide in-plane-momentum matching for SPP excitation by free-space photons if $|k_{\text{SPP}} - g_l| \leq \omega/c$. Free-space photons with in-plane wavevector component $\mathbf{k}_{\parallel} = (k_x, 0)$ can excite SPPs if the momentum is

matched through some sinusoid $l$ to a coupled mode $j$ with significant SPP character $v_{j,0}$. We account for photon–SPP momentum matching again by considering the effect of grating components $l$ only in first order. We generate $N$ copies of this dispersion by shifting the wavevector value to $k_{i,l} = k_{\text{SPP},i} - g_l$ for all $l \in \{1,2,\ldots,N\}$. These are the $k$-values from which grating component $l$ could enable SPP incoupling. Then we copy and mirror the entire dispersion in the $(k=0)$-axis, realizing that the entire problem is symmetric under inversion of the propagation direction of the modes. We thus obtain $2N$ copies of our calculated dispersion, some of which may entirely fall outside the experimental range of wavevectors and energies. We consider that at each point $(k_{i,l}, E_{i,j})$ or $(-k_{i,l}, E_{i,j})$, with $i \in \{1,2,\ldots,M\}$, $j \in \{1,2,\ldots,N\}$, and $l \in \{1,2,\ldots,N\}$, the coupling to grating SPPs is proportional to $\Gamma_l v_{i,j,0}^2$. This reflects that, for first-order coupling, the magnitude of the admixture is proportional to the SPP character of the coupled mode. We thus obtain a model function for the incoupling $V$ as a function of the photon in-plane wavevector component $k_x$ and energy $\hbar\omega$ of

$$V(k_x, \hbar\omega) = \sum_{i=1}^{M}\sum_{j=1}^{N}\sum_{l=1}^{N} \Gamma_l v_{i,j,0}^2 \delta(k_x \pm k_{i,l})\delta(\hbar\omega - E_{i,j}), \qquad (15)$$

where $\delta(x)$ is the Kronecker delta function. Finally, we broaden $V$ by convolution with a function

$$P(k_x) = \operatorname{sinc}^2(k_x d/2) \qquad (16)$$

in the $k_x$-direction to account for the finite length $d = 9$ μm of our gratings, and with a Gaussian function $Q(\hbar\omega)$ with a variance of $\sigma^2 = (15\text{ meV})^2$ in the $\hbar\omega$-direction to match the experimental broadening, due to a combination of finite instrumental resolution, the finite range of $k_y$ values for reflected photons, and losses. The convolved function $(V * P * Q)(k_x, \hbar\omega)$ is plotted in Figs. 1c,f,i of the main text.

**Method references**

39. Moreno, V., Román, J. F. & Salgueiro, J. R. High efficiency diffractive lenses: deduction of kinoform profile. *Am. J. Phys.* **65**, 556-562 (1997).

40. McPeak, K. M., Jayanti, S. V., Kress, S. J. P., Meyer, S., Iotti, S., Rossinelli, A. & Norris, D. J. Plasmonic films can easily be better: rules and recipes. *ACS Photonics* **2**, 326-333 (2015).

|  | **Parameters** | | | | | | | | | | **Height profile** |
|---|---|---|---|---|---|---|---|---|---|---|---|
| **Figure** | $A_1$ (nm) | $A_2$ (nm) | $A_3$ (nm) | $\Lambda_1$ (nm) | $\Lambda_2$ (nm) | $\Lambda_3$ (nm) | $\varphi_1$ (deg) | $\varphi_2$ (deg) | $\varphi_3$ (deg) | $\Delta$ (nm) | $f(x)$ |
| Fig. 1a | 25.0 | – | – | 600 | – | – | 180 | – | – | 35.0 | |
| Fig. 1d | 18.1 | 14.3 | – | 600 | 475 | – | 0 | 0 | – | 42.3 | |
| Fig. 1g | 18.4 | 7.0 | 6.4 | 600 | 230 | 210 | 0 | 0 | 0 | 41.8 | |
| Ext. Data Fig. 4 | 19.3 | 9.6 | – | 620 | 310 | – | 180 | 180 / 0 | – | 24.4 / 38.9 | |
| Ext. Data Fig. 5 | 18.5 | 0 to 25.1 | – | 620 | 230 | – | 0 | 0 | – | 28.0 to 53.1 | $\sum_i A_i \cos(g_i x + \varphi_i) - \Delta$ |
| Fig. 4b | 18.2 | 15.2 | 13.0 | 620 | 520 | 445 | 0 | 0 | 0 | 56.5 | |
| Ext. Data Fig. 8a | 25.0 | – | – | 400 | – | – | 180 | – | – | 35.0 | |
| Ext. Data Fig. 8c | 14.1 | 11.8 | 10.1 | 414 | 347 | 297 | 0 | 0 | 0 | 45.8 | |

| **Figure** | $i$ | $A_i$ (nm) | $\Lambda_i$ (nm) | $\varphi_i$ (deg) | $\theta_{1,7}$ (deg) | $\theta_{2,8}$ (deg) | $\theta_{3,9}$ (deg) | $\theta_{4,10}$ (deg) | $\theta_{5,11}$ (deg) | $\theta_{6,12}$ (deg) | $\Delta$ (nm) | $f(x,y)$ |
|---|---|---|---|---|---|---|---|---|---|---|---|---|
| Fig. 2a | 1,2 | 17.5 | 600 | 0 | 0 | -10 | – | – | – | – | 45.0 | |
| Fig. 2b | 1,2 | 17.5 | 600 | 0 | 0 | -40 | – | – | – | – | 45.0 | |
| Fig. 3a | 1,2,3 | 15.6 | 600 | 0 | 0 | 60 | 120 | – | – | – | 56.7 | |
| Fig. 3d | 1–6 | 10.0 | 600 | 0 | 0 | 30 | 60 | 90 | 120 | 150 | 70.0 | $\sum_i A_i \cos[g_i(x\cos\theta_i + y\sin\theta_i) + \varphi_i] - \Delta$ |
| Fig. 4e | 1–6 | 5.6 | 700 | 0 | 0 | 30 | 60 | 90 | 120 | 150 | 77.3 | |
| Fig. 4e | 7–12 | 5.6 | 308 | 0 | 0 | 30 | 60 | 90 | 120 | 150 | 77.3 | |
| Ext. Data Fig. 9 | 1–6 | 11.1 | 615 | 0 | 0 | 30 | 60 | 90 | 120 | 150 | 76.7 | |

| **Figure** | $i$ | $A_1$ (nm) | $A_2$ (nm) | $\Lambda_1$ (nm) | $\Lambda_2$ (nm) | $\varphi_1$ (deg) | $\varphi_2$ (deg) | $\Delta$ (nm) | $\mathbf{r}_1$ (nm) | $\mathbf{r}_2$ (nm) | $f(r,\theta)$ |
|---|---|---|---|---|---|---|---|---|---|---|---|
| Ext. Data Fig. 7a | 1 | 35.0 | – | 600 | – | 180 | – | 45 | **0** | – | $\sum_i A_i \cos(g_i|\mathbf{r} - \mathbf{r}_i| + \varphi_i) - \Delta$ |
| Ext. Data Fig. 7b | 1,2 | 17.5 | 17.5 | 600 | 600 | 180 | 0 | 45 | $-150\,\hat{\mathbf{y}}$ | $150\,\hat{\mathbf{y}}$ | |

| **Figure** | $A$ (nm) | $L$ (nm) | $\Delta$ (nm) | $f(r)$ |
|---|---|---|---|---|
| Fig. 2e | 35.0 | 1581 | 45.0 | $A\sin\left[\pi\left(\dfrac{r}{L}\right)^2\right] - \Delta$ |

**Extended Data Table 1 | Design parameters for Fourier surfaces.** Design parameters for all Fourier surfaces demonstrated in this work. The functions are defined for the design to be patterned in the polymer surface, where $x$ and $y$ lie in plane and $z$ is perpendicular (pointing away from the substrate). A right-handed coordinate system is used with the origin placed in the middle of the pattern in both the $x$ and $y$ directions. In these formulas, the height of the surface is defined relative to the unpatterned flat surface where $z = 0$. All $A_i$ and $\Delta_i$ ($\Lambda_i$) have been rounded to the nearest 0.1 (1.0) nm. Analysis of the measured topographies for templated Ag gratings shows that the $\Lambda_i$ are consistently ~2% larger than the design value (Extended Data Fig. 2), attributed to a distance miscalibration in the thermal scanning probe. See Methods.

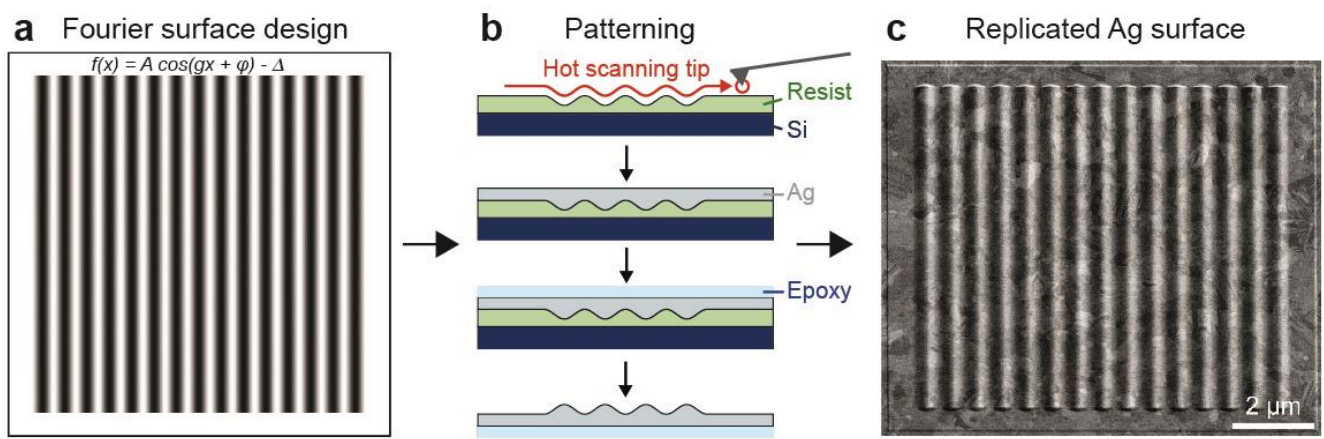

**Extended Data Figure 1 | Design and fabrication of Fourier surfaces. a**, Design of a Fourier surface. The analytical formula for the desired surface profile (here, a single 1D sinusoid) is converted into a grayscale bitmap. Each 10×10 nm$^2$ pixel has a depth level between 0 and 255 (8-bit). The bitmap contains the sinusoidal function in the horizontal direction within the white border, which is constant along the vertical direction. The pixels in the white border are set to the minimum depth level. **b**, Process flow showing our patterning steps for Ag Fourier surfaces: (i) The hot scanning tip is used to create a single sinusoid in the polymer resist, (ii) An optically thick (>500 nm) Ag layer is thermally evaporated onto the polymer, (iii) A glass microscope slide is affixed to the back of the Ag layer using UV-curable epoxy, and (iv) the glass/epoxy/Ag stack is stripped off the polymer film. Alternative fabrication pathways for transferring the Fourier surface pattern to other materials are presented in Methods. **c**, SEM (30° tilt) of single 1D sinusoidal Fourier surface transferred to Ag via templating. The initial analytical design is replicated accurately in the final Ag surface.

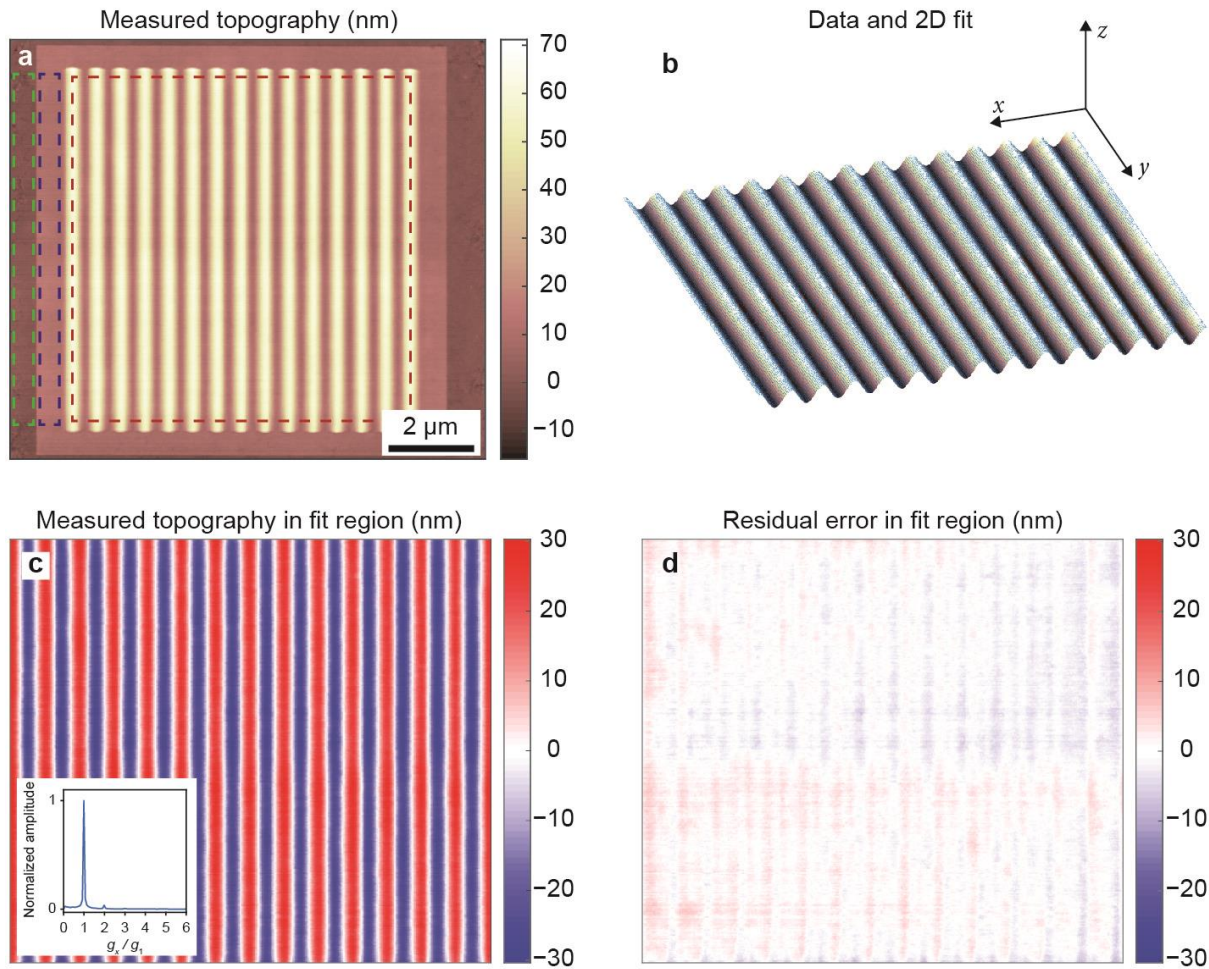

**Extended Data Figure 2 | Topography characterization. a**, AFM micrograph of the measured topography for a single-sinusoidal Ag grating. The RMS roughness of the unpatterned flat Ag film is 1.6 nm, extracted from the area indicated by the green dashed box. The RMS roughness of the patterned flat Ag film is 1.3 nm, extracted from the area indicated by the blue dashed box. The area indicated by the red dashed box is used for fitting and analysis of the surface profile. **b**, 2D fit of a sinusoidal function (yellow/brown surface) to topography data (blue dots) from the region indicated in the red dashed box in **a**. The amplitude of the fitted function is $A_1 = 25.5$ nm (2% larger than design value) with a period of $\Lambda = 610$ nm (1.7% larger than design value). Such horizontal errors were consistent over many samples and attributed to a distance miscalibration in the thermal scanning probe. The RMS error between the 2D design function and measured topography was found to be 1.8 nm after this horizontal error was taken into account. **c**, Measured topography of the structure in **a**, plotted only for the fit region (red dashed box in **a**), scaled from the minimum depth value to the maximum depth value and centered at zero. The inset shows a line cut (along $g_x$ at $g_y = 0$, where $g_x$ and $g_y$ are the projections of **g** along $x$ and $y$, respectively) from the 2D Fourier transform of the measured topography in the fit region, normalized to the peak value at $g_1$. The second harmonic at $\frac{g_x}{g_1} = 2$ is barely visible and has an amplitude of 3.5% of the peak at $\frac{g_x}{g_1} = 1$, corresponding to a real-space amplitude of 0.9 nm. **d**, Residual error between the data and the fitted function, plotted for the fit region as in **c**. For comparison, the data is scaled over the same range as in **c**, centered at 0.

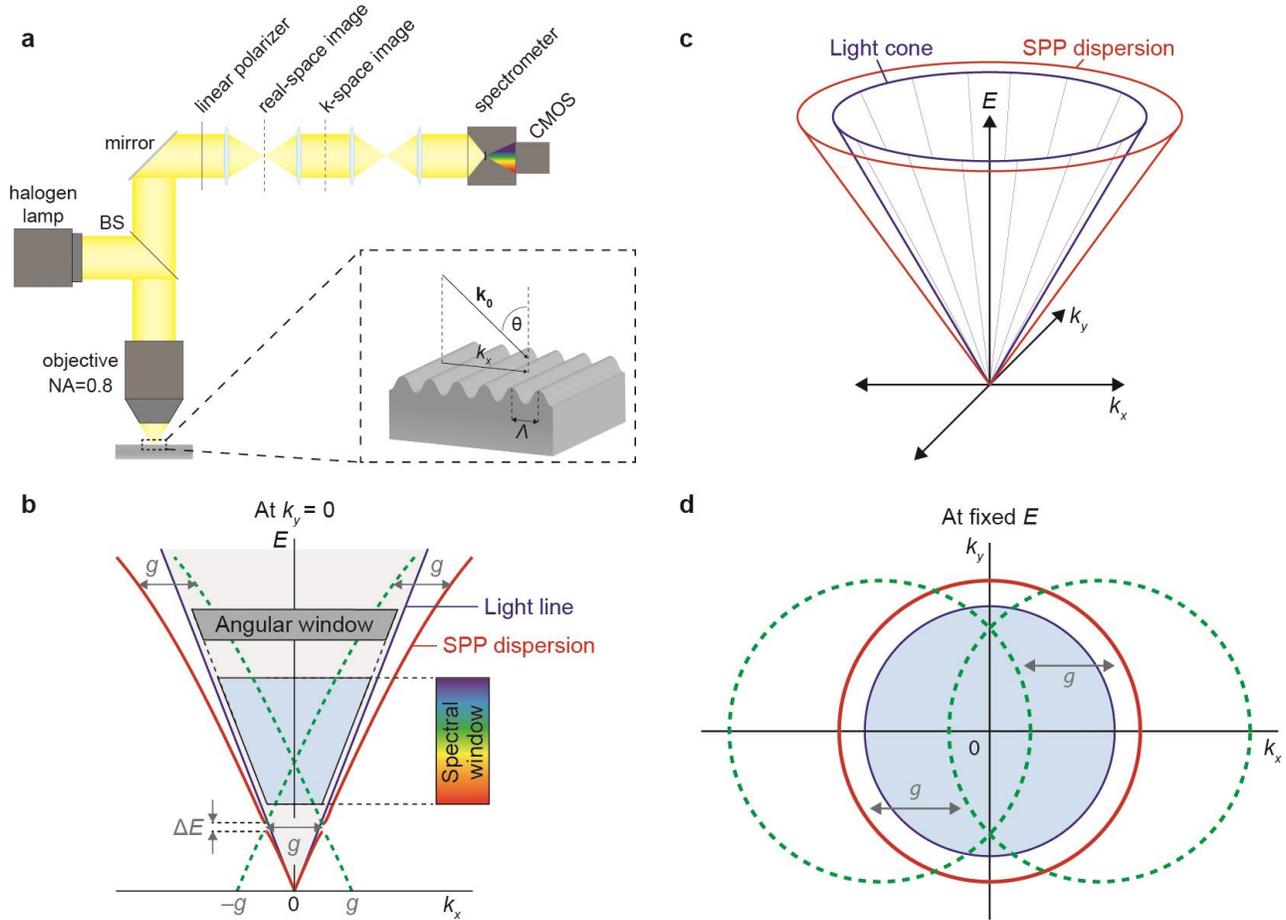

**Extended Data Figure 3 | Optical measurement of plasmonic Fourier surfaces. a**, Schematic of the optical setup used for $k$-space reflectivity measurements. Further details are in Methods. Inset: vector diagram of light with wavevector $\mathbf{k_0}$ incident at angle $\theta$ on a Fourier surface pattern with period $\Lambda$. **b**, Schematic of the dispersion diagram (energy *versus* in-plane wavevector, $k_x$) for free-space photons incident on a 1D sinusoidal grating with $k_y = 0$ (as in Fig. 1 in the main text). By tuning $\theta$, photons have access to the shaded region inside the light lines (solid blue lines). The red lines show the SPP dispersion, $k_{\text{SPP}}$. Dashed green curves indicate the SPP dispersion displaced by the grating spatial frequency $g$. Inside the light line, these curves represent where free-space photons can couple to SPPs, and vice-versa (that is, where $k_x \pm g = k_{\text{SPP}}$). A stopband of width $\Delta E$ opens when counter-propagating SPPs are coupled by $g$. The blue trapezoidal region depicts the experimentally accessible area on the dispersion diagram, limited by the spectral window of the spectrometer along $E$, and the angular window of reflected light collected by the microscope objective along $k_x$. **c**, Schematic of the dispersion diagram for free-space photons incident on a surface, plotted for both in-plane wavevectors, $k_x$ and $k_y$. The light line and SPP dispersion in **b** are both cones (blue and red lines, respectively). **d**, A slice through the dispersion diagram in **c** at fixed energy. Free-space photons incident on a surface can have wavevectors inside the light cone (blue shaded region). The SPP dispersion is the larger red circle. Dashed green circles show solutions to $\mathbf{k_\parallel} \pm \mathbf{g} = \mathbf{k_{\text{SPP}}}$. In this example, $\mathbf{g} = g\,\hat{\mathbf{x}}$.

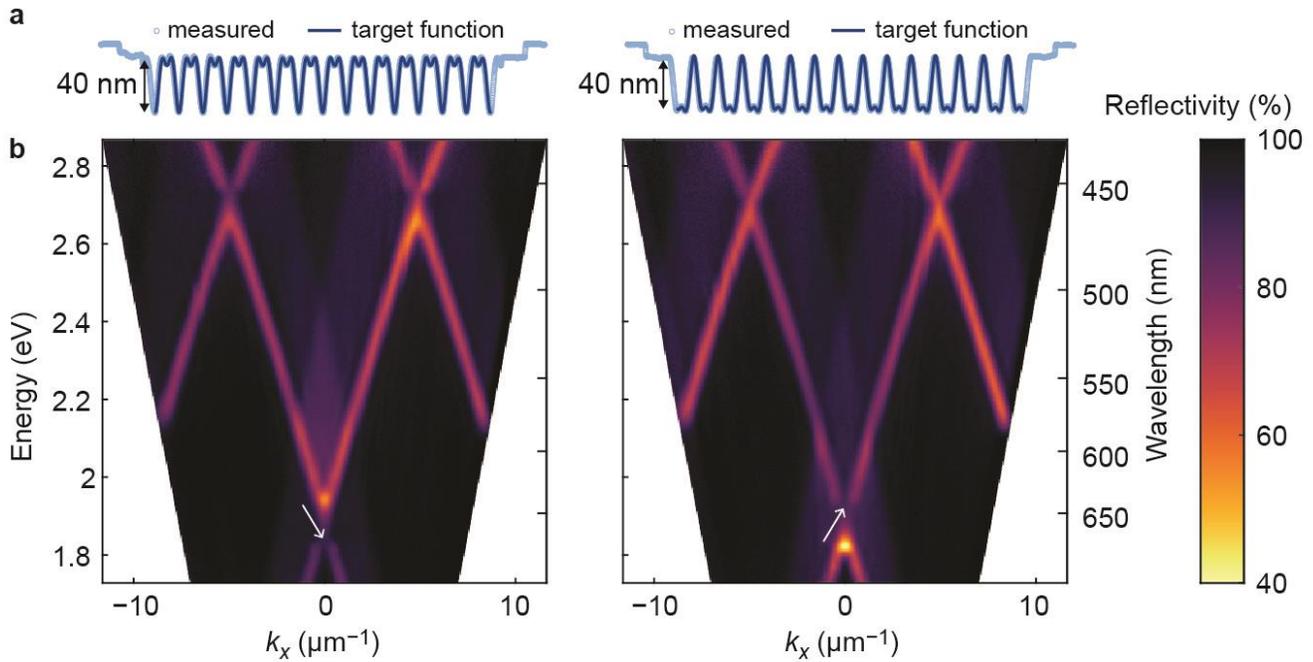

**Extended Data Figure 4 | Control of 'dark' band edges in two-component 1D sinusoidal gratings. a**, Comparisons of the measured (light blue points) and targeted surface topographies (dark blue lines) in the polymer surface, measured during patterning. Scan lengths are 11.5 μm. The left grating has the height profile $f(x) = A_1 \cos(gx + \pi) + \frac{A_1}{2}\cos(2gx + \varphi_2) - \Delta$ with $\varphi_2 = \pi$. The grating on the right has the same $f(x)$ except $\varphi_2 = 0$. **b**, The measured reflectivity in $k$-space (as in Fig. 1 in the main text) for Ag gratings templated from the structures in **a**. In both the left and right gratings, a stopband opens near 1.9 eV, but the choice of phase can control whether an optically dark state exists at the lower (left) or upper (right) band edge. The band edge with the optically dark state is marked with white arrows. For all structural design parameters, see Extended Data Table 1.

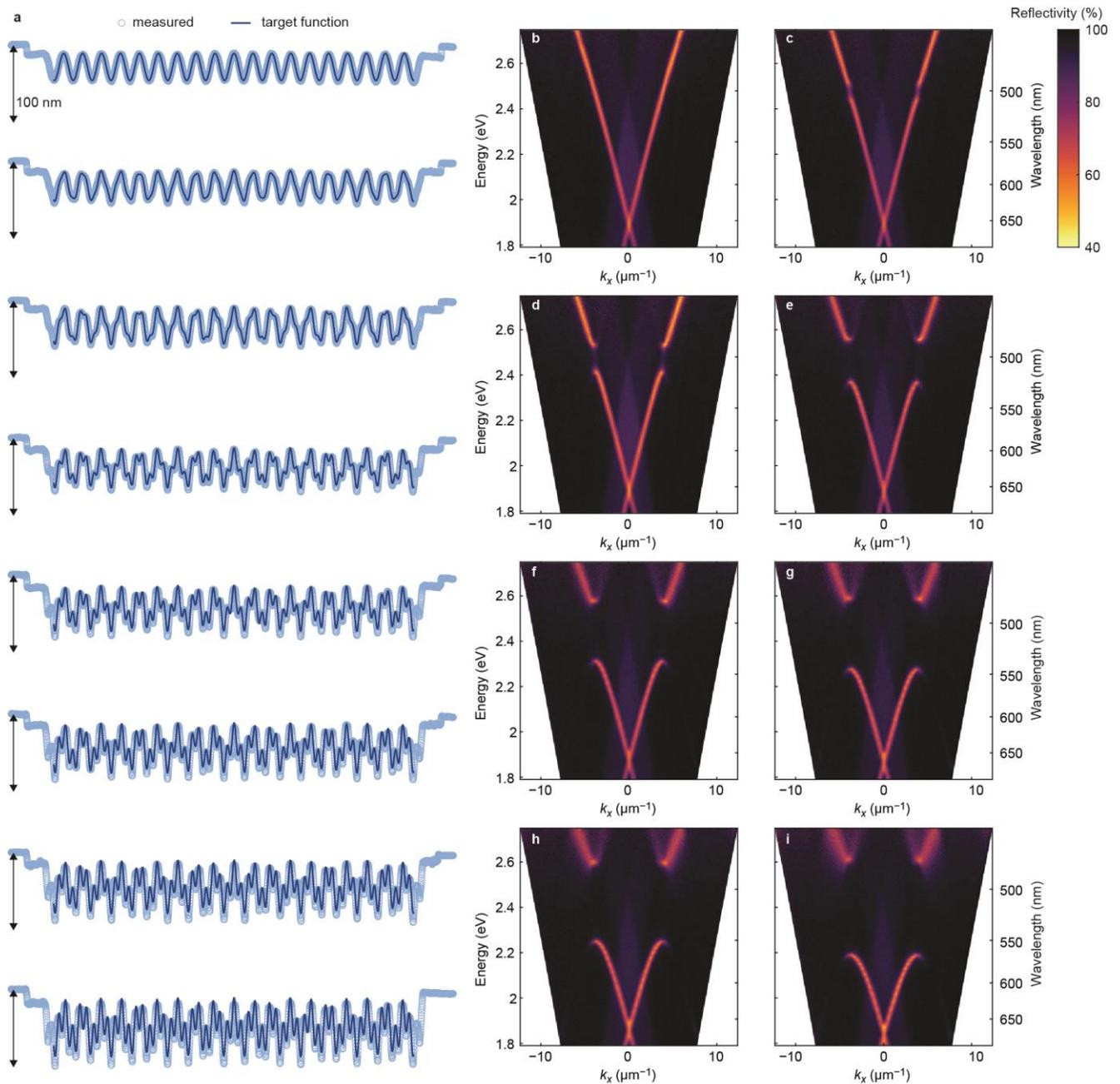

**Extended Data Figure 5 | Control of stopband width. a**, Comparisons of the measured (light blue points) and targeted surface topographies (dark blue lines) in the polymer surfaces, measured during patterning, for structures exhibiting a single stopband. Scan lengths are 14.5 µm and the vertical scale bar is 100 nm for all scans. From top to bottom: a series of two-component 1D sinusoidal gratings, where $A_1$ = 18 nm, $\Lambda_1$ = 620 nm, $A_2$ is varied, and $\Lambda_2$ = 230 nm. $A_2$ has values of 0, 2.5, 5, 10, 15, 18, 20, 25.1 nm. **b-i**, Measured plasmonic dispersion diagrams for Ag gratings templated from the profiles in **a**, from top to bottom, respectively. The width of the stopband increases because $A_2$ is the amplitude of the Fourier component responsible for creating the plasmonic stopband. For all structural design parameters, see Extended Data Table 1.

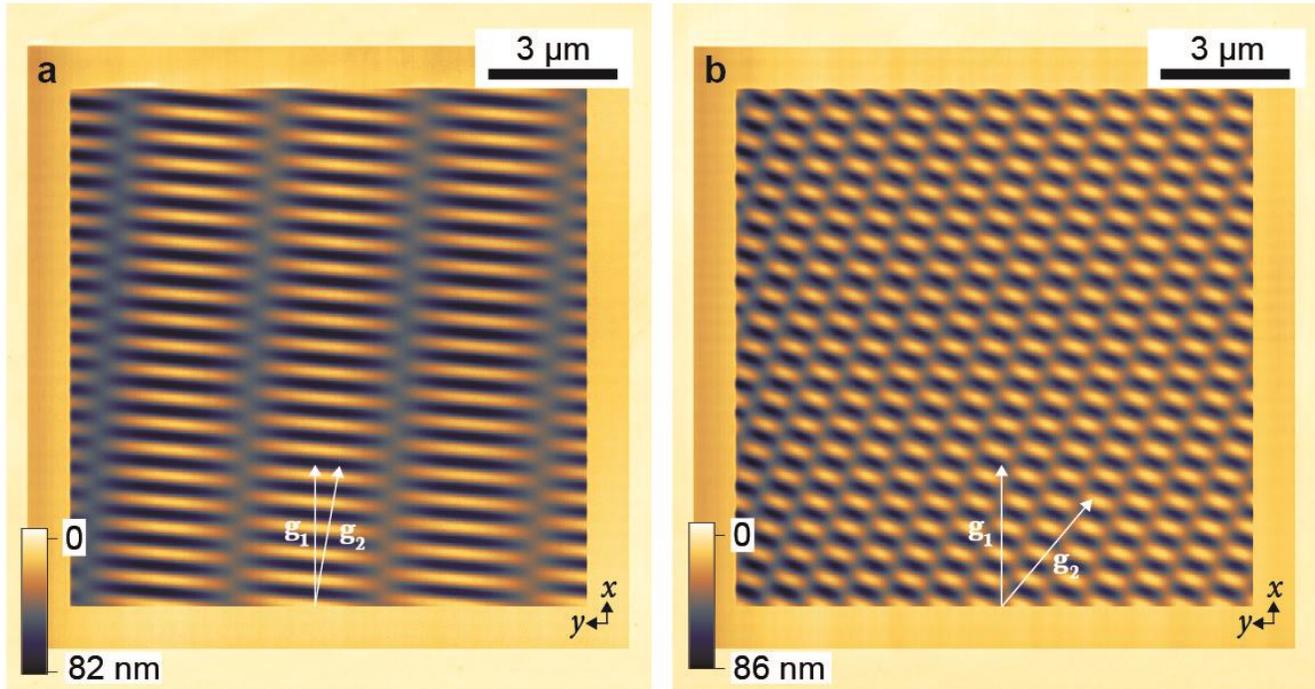

**Extended Data Figure 6 | Spatial-frequency vectors for 2D Fourier surface patterns. a**, Measured topography (obtained during patterning) for the polymer film (PMMA/MA, see Methods) used to template the structure in Fig. 2a of the main text. The two spatial-frequency vectors $\mathbf{g_1}$ and $\mathbf{g_2}$ that define the surface profile are overlaid on the pattern. Here, $\mathbf{g_1}$ and $\mathbf{g_2}$ have the same magnitude $g_1 = g_2 = \frac{2\pi}{600\,\text{nm}}$, and $\mathbf{g_2}$ is rotated $-10°$ from $\mathbf{g_1}$, where $\mathbf{g_1}$ lies along $\hat{\mathbf{x}}$. **b**, As in **a**, but the template corresponding to the structure in Fig. 2b of the main text. Again, $\mathbf{g_1}$ and $\mathbf{g_2}$ have the same magnitude $g_1 = g_2 = \frac{2\pi}{600\,\text{nm}}$, but now $\mathbf{g_2}$ is rotated $-40°$ from $\mathbf{g_1}$, where $\mathbf{g_1}$ lies along $\hat{\mathbf{x}}$. See Extended Data Table 1 for further details.

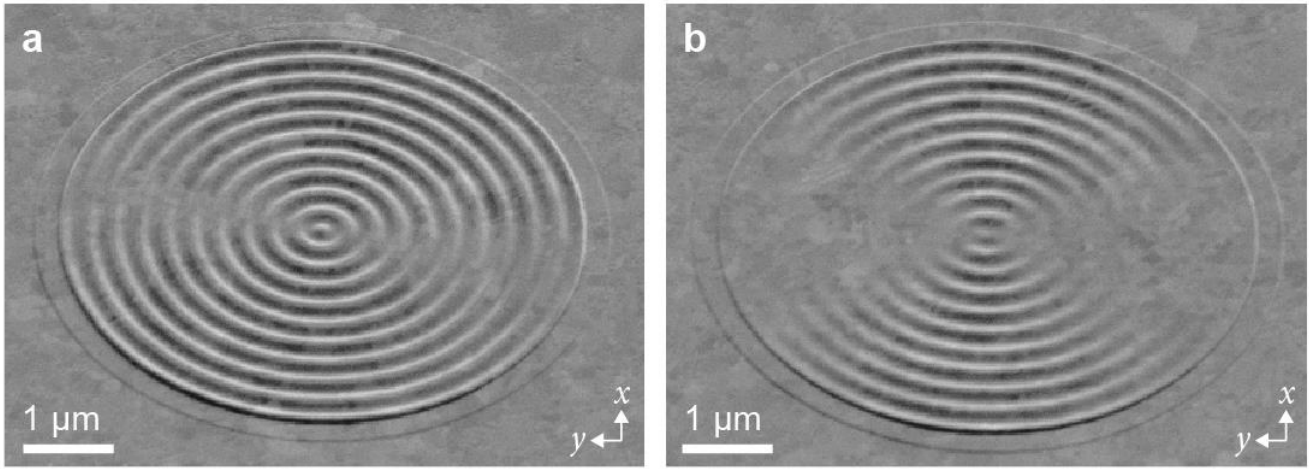

**Extended Data Figure 7 | 2D Fourier surfaces with circular basis functions. a**, SEM (45° tilt) of a circular sinusoidal Ag grating with $\Lambda$ = 600 nm. **b**, SEM (45° tilt) of two superimposed circular sinusoidal gratings, as in **a**, each with $\Lambda$ = 600 nm. The center of one grating is translated +150 nm and the other −150nm in $\hat{y}$ from the origin in the middle of the pattern. The spatial interference results in a moiré pattern with broken circular symmetry. For all structural design parameters, see Extended Data Table 1.

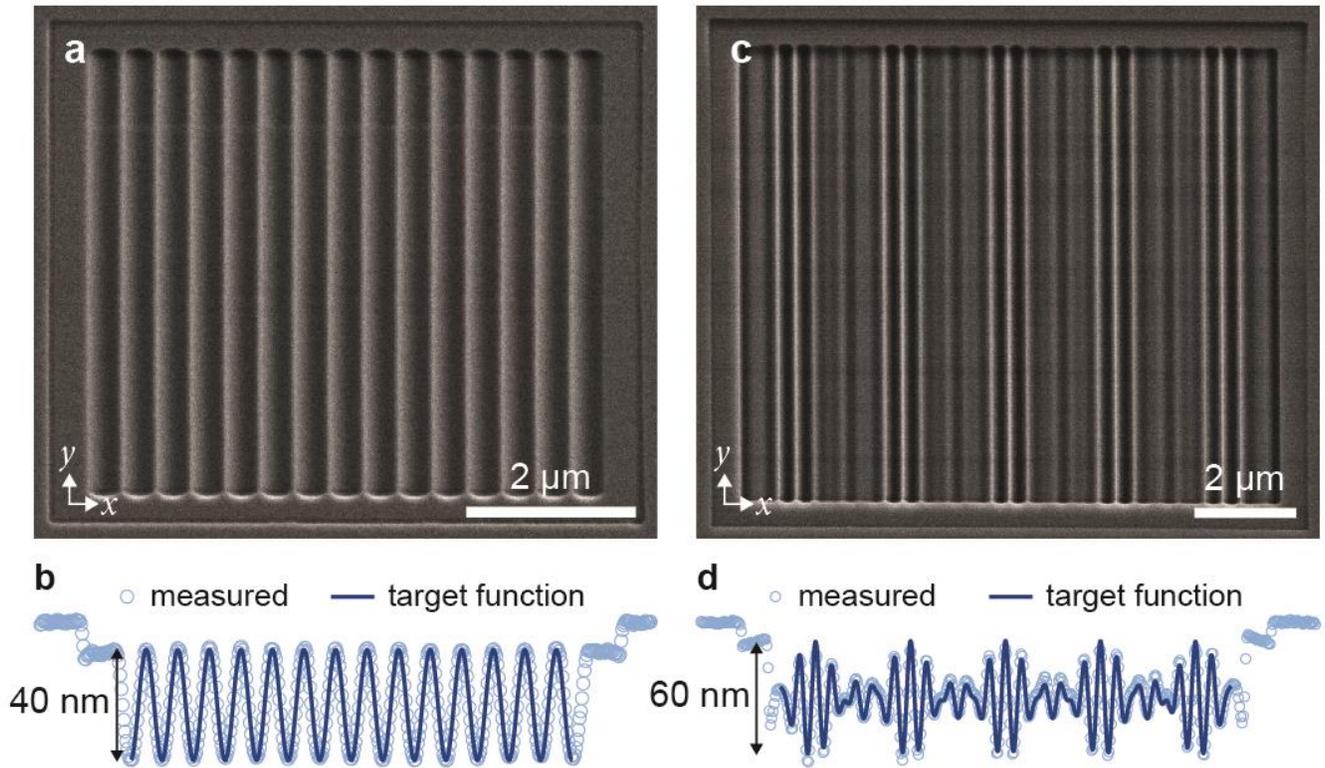

**Extended Data Figure 8 | SiN$_x$ Fourier surfaces. a**, SEM (30° tilt) of a single 1D sinusoid in SiN$_x$, transferred via reactive ion etching (RIE) (see Methods for details). **b**, Comparison of the measured (AFM) and targeted surface topography (accounting for a slight distance miscalibration in the thermal scanning probe). Scan length is 11.3 µm. The final profile in SiN$_x$ has a measured RMS error of 2.5 nm using the same methodology as in Extended Data Fig. 2. **c**, As in **a**, but for a three-component 1D SiN$_x$ grating. **d**, As in **b**, but for the structure in **c**. Scan length is 14.8 µm. The final profile in SiN$_x$ has a measured RMS error of 3.9 nm using the same methodology as in Extended Data Fig. 2. For all structural design parameters, see Extended Data Table 1.

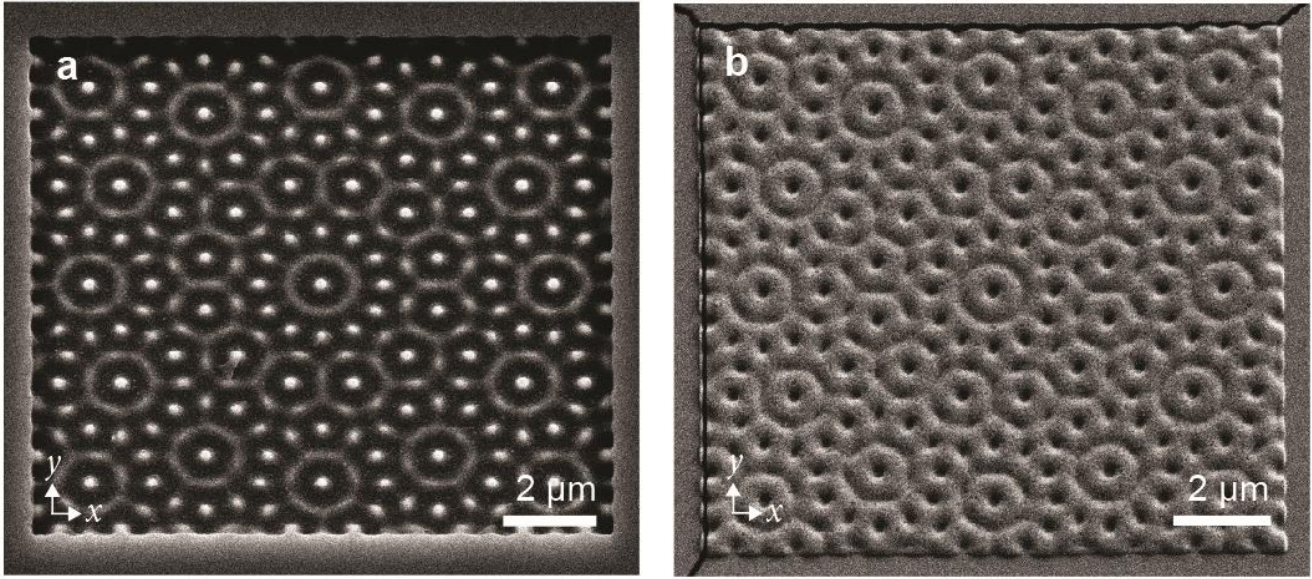

**Extended Data Figure 9 | TiO$_2$ Fourier surface from a patterned Si template. a**, SEM (30° tilt) of a 12-fold rotationally symmetric quasicrystal, as in Fig. 3d of the main text, transferred from the patterned polymer to Si via inductively coupled plasma (ICP) etching (see Methods for details). **b**, SEM (30° tilt) of the pattern in **a** transferred to a TiO$_2$ thin-film via template-stripping (see Methods for details). For all structural design parameters, see Extended Data Table 1.

35